\documentclass[a4paper,12pt]{article}
\usepackage{amsmath,amssymb}
\begin{document}
\thispagestyle{empty}

\begin{center}
{\bf LECTURES ON INTEGRABLE HIERARCHIES \\ AND VERTEX OPERATORS}
\end{center}

\vspace{0cm}

\begin{center}
{\sc A.\,A.\,Vladimirov} \\[.1cm]
{\em BLTP, JINR, Dubna, Moscow region 141980, Russia \\
alvladim@thsun1.jinr.ru}
\end{center}

\vspace{1.2cm}

\begin{center}
\large\textbf{Contents}
\end{center}

\vspace{.1cm}

\noindent
I. \ \,INTEGRABLE HIERARCHIES

Scalar Ansatz (KP hierarchy) \dotfill \ \ 2

Fermionic Fock space \dotfill \ \ 7

Fermi-Bose correspondence \dotfill \ 10

KP hierarchy via free fermions \dotfill \ 14

\noindent
II. VERTEX ALGEBRAS

Formal distributions and locality \dotfill \ 15

Operator product expansion \dotfill \ 17

Vertex algebras \dotfill \ 19

Free fermions \dotfill \ 20

Virasoro algebra in KdV \dotfill \ 23

\vspace{1.2cm}

\begin{center}
\large\textbf{Abstract}
\end{center}

This is a slightly extended
write-up of lectures given at the 2004 DIAS-TH Winter school
on theoretical physics in Dubna, Russia. The exposition is intended for
undergraduate students. The two parts of this course are linked together
by the concept of vertex operators, which naturally arise, via Fermi-Bose
correspondence, in the context of integrable hierarchies, and later play
a central role in vertex algebras, under the name of the state-field map.
Also, the final section illustrates another deep connection between
integrable (here KdV) hierarchies and vertex (here Virasoro) algebras.
Last updates and some related material can be found on my site
http://thsun1.jinr.ru/${}^\sim$alvladim/qft.html

The lectures do not include any original results of the author, and are
mostly compiled from well-known textbooks and introductory papers.
Main references to be cited here are:

\vspace{.2cm}

T.\,Miwa, M.\,Jimbo and E.\,Date, \,\textit{Solitons}, 2000.

V.\,Kac, \ \textit{Infinite Dimensional Lie Algebras}, 1990.

V.\,Kac, \,\textit{Vertex Algebras for Beginners}, 1998.

H.\,Aratyn, hep-th/9503211.

A.\,Matsuo and K.\,Nagatomo, hep-th/9706118.

E.\,Witten, Comm.Math.Phys. 114 (1988) 1.

\newpage

\begin{center}
\large\textbf{I. \ INTEGRABLE HIERARCHIES}
\end{center}

\vspace{0cm}

\begin{center}
\large\textbf{Scalar Ansatz (KP hierarchy)}
\end{center}

\vspace{.1cm}

Probably, the most efficient way to deal with integrable
nonlinear equations is the Lax representation
\begin{equation}
\label{Lax}
\partial_t L = [M,L]\,.
\end{equation}
This readily produces a formally infinite set of conserved quantities
$I_n\sim\text{tr}\,L^n$\,:
\begin{equation}
\label{trLn}
(1) \ \ \ \Rightarrow \ \ \ \partial_t L^n = [M,L^n] \ \
\Rightarrow \ \ \partial_t\,\text{tr}L^n = \text{tr}[M,L^n] = 0\,,
\end{equation}
and enables us to view the $t$-evolution of $L$ as a similarity transformation,
\begin{equation}
\label{TLT}
\partial_t T = MT\,, \ \ \ \partial_t T^{-1} = -T^{-1}M \ \ \
\Rightarrow \ \ \ L(t) = TL(0)T^{-1}\,,
\end{equation}
which implies isospectrality of the corresponding eigenvalue problem:
\begin{equation}
\label{eigen}
L(t)\psi(t) = z\,\psi(t) \ \ \ \ \Rightarrow \ \ \ \partial_t z = 0\,,\ \ \ \
\psi(t) = T\psi(0)\,,  \ \ \ \partial_t \psi = M \psi\,.
\end{equation}

To find a Lax (or $L-\!M$) pair for a given equation is an extremely difficult
(and generally unsolvable) task. If one, on the contrary, starts with some
reasonable Ansatz for $L$, it appears fairly possible to discover an infinite
set $\{M_n\}$ of appropriate $M$-operators. If, moreover, the flows given by
the corresponding equations (\ref{Lax}) commute (so that the $t_n$-evolutions
may be considered as independent, and the $t_n$ themselves -- as a new set of
coordinates on the phase space), the whole set of equations (\ref{Lax}) with
$M_n$ is called an (integrable) hierarchy.

In practice, integrable hierarchies are highly symmetric, infinite sets of
nonlinear evolution equations of the Lax type for infinitely many functions
$u_i$ of infinitely many variables $t_n, \ n=1,2,\,\ldots $\,.
We use the notation $x\equiv t_1, \ \partial\equiv\partial_x\equiv\partial_1,
\ \partial_n\equiv\frac{\partial}{\partial t_n}$. When needed, we will
specially indicate the operator nature of a derivative,
e.g., $\partial_n\circ B = \partial_n B  + B\,\partial_n$\,.
Formal sums of the type
\begin{equation}
\label{pseudo}
P(x,\partial) = \sum_{m}a_m(x)\,\partial^m\,, \ \ \ \ \ \
a_m(x)=0 \ \ \text{for} \ \ m\gg 0
\end{equation}
are called pseudodifferential operators (in a variable $x$. Other possible
arguments of $a_m$ are treated as parameters).
As usual,
\begin{equation}
\label{+-}
P_+ = \sum_{m\geqslant0}a_m(x)\,\partial^m\,, \ \ \
P_- = P - P_+ = \sum_{m<0}a_m(x)\,\partial^m\,, \ \ \
\text{Res}P = a_{-1}(x)\,,
\end{equation}
so that $P_+$ is a proper differential part of $P$.
Operator $\partial^{-1}$ is defined by
$\partial\,\partial^{-1}=\partial^{-1}\,\partial=1$\,. For any integer $m$,
the following holds:
\begin{equation}
\label{diff}
\partial^m\circ f = \sum_{k\geqslant0}\binom{m}{k}\partial^k\!f\,\partial^{m-k}\,.
\end{equation}
With this rule, pseudodifferential operators form an associative algebra.
This is shown straightforwardly, as well as the following useful relation:
\begin{equation}
\label{res}
\text{Res}[P,Q] = \partial(\ldots)\,.
\end{equation}

The Kadomtsev-Petviashvili (KP) hierarchy is generated by the following
(scalar) Ansatz for the Lax operator:
\begin{equation}
\label{scalar}
L = \partial + \sum_{i=1}^{\infty}u_i\partial^{-i}\,.
\end{equation}
Unknown functions $u_i$ may depend on all arguments ${t_n}$\,, or on
a part of them. We use
\begin{equation}
\label{Bn}
B_n \doteq (L^n)_+ \equiv L^n_+ \ \ \ \ \
(B_1 = \partial\,, \ \ B_2 = \partial^2 +2u_1\,, \ \
B_3 = \partial^3 + 3\partial_x\!u_1 + 3u_1\partial +3u_2\,, \ \ldots)
\end{equation}
to formulate the infinite system of equations of the KP hierarchy:
\begin{equation}
\label{main}
\partial_n L = [B_n,L] \ \ \ \text{or} \ \ \
[\partial_n - B_n\,, L] = 0 \ \ \ \ \ (n = 1,2,\ldots)
\end{equation}
These are nonlinear equations in $u_i$. Their integrability
is based on the Lax form of (\ref{main}) and on the following
``zero-curvature" property,
\begin{equation}
\label{0curv}
[\partial_m - B_m\,,\partial_n - B_n] =
\partial_n B_m - \partial_m B_n  + [B_m,B_n] = 0\,.
\end{equation}
From this, one easily deduces that the flows generated by different
$B_n$ commute,
\begin{equation}
\label{comm}
\partial_n[B_m,L] = \partial_m[B_n,L]\,.
\end{equation}

To derive (\ref{0curv}) itself, we observe that (\ref{main}) entails
$\partial_n L^m = [B_n,L^m]$\,, that $[B_n,L^m]=-[L^n_-,L^m]$\,, and then
do the following calculations:
\begin{multline}
\label{der0}
\partial_n L^m - \partial_m L^n = [B_n,L^m] - [B_m,L^n]
= [B_n,L^m] + [L_-^m,L^n] \\
= [B_n,B_m] + [B_n,L^m_-] + [L_-^m,L^n] = [B_n,B_m] + [L^m_-,L_-^n]\,.
\end{multline}
A positive part of this equality is exactly (\ref{0curv}).

The KP equation for $u_1=u(x,y,t), \ y=t_2, \ t=t_3$\,,
\begin{equation}
\label{KP}
3u_{yy} + \partial_x(-4u_t + u_{xxx} +12\,u u_x) = 0\,,
\end{equation}
which gave its name to the whole hierarchy, is extracted either from
(\ref{0curv}) for $m=2, n=3$\,, or from (\ref{main}) with $n=2,3$\,.

The main equations (\ref{main}) can be viewed as the compatibility
condition of the linear system
\begin{equation}
\label{Baker}
L\psi = z\psi\,, \ \ \ \ \ \ \partial_n\psi = B_n\psi\,.
\end{equation}
Here $\psi(t,z)\equiv \psi(\{t_n\},z)$ is known as Baker's function.
It admits an expansion
\begin{equation}
\label{lnpsi}
\ln\psi = \sum_{n=1}^{\infty}z^n t_n + \sum_{i=1}^{\infty}z^{-i}\psi_i(t)\,.
\end{equation}
To show this, we note that $\partial$, and $B_m$ in general,
can be re-expanded in powers of $L$\,:
\begin{gather}
\label{reexpand}
\partial = L + \sum_{i=1}^{\infty}\sigma_{i}^{(1)}L^{-i}\,, \\
\label{Bm}
B_m = L^m + \sum_{i=1}^{\infty}\sigma_{i}^{(m)}L^{-i}\,,
\end{gather}
all $\sigma_{i}^{(m)}$ being expressible in $\sigma_{i}^{(1)}$\,.
From (\ref{Baker}) we see that $L^m\psi=z^m\psi$ and
\begin{equation}
\label{dlnpsi}
\partial_m\psi = (z^m + \sum_{i=1}^{\infty}z^{-i}\sigma_{i}^{(m)})\psi\,,
\end{equation}
which agrees with (\ref{lnpsi}) if we fix
\begin{equation}
\label{dpsi}
\partial_m\psi_n =\sigma_{n}^{(m)}\,.
\end{equation}
Further, we observe that
\begin{equation}
\label{set1}
\partial_k\sigma_{n}^{(m)} = \partial_m\sigma_{n}^{(k)}
= \partial_k\partial_m\psi_n\,,
\end{equation}
and, as a corollary, $\partial_k\sigma_{n}^{(1)} = \partial(\ldots)$\,,
i.e., $\sigma_{n}^{(1)}$ are conserved currents. Another set of such
currents is provided by $\sigma_{1}^{(m)}=-\text{Res}L^m$
(due to (\ref{Bm}))\,:
\begin{equation}
\label{set2}
\partial_k\sigma_{1}^{(m)} = -\text{Res}\partial_k L^m
= -\text{Res}[B_k,L^m] = \partial(\ldots)\,.
\end{equation}
These two sets are not independent: it can be shown that
\begin{equation}
\label{recurs1}
\sigma_{1}^{(n)} - n\sigma_{n}^{(1)}
=\sum_{m=1}^{n-1}\partial_m\sigma_{n-m}^{(1)}\,.
\end{equation}

Let us briefly describe the derivation. Using in $L^{m+1}=LL^m$
eq.\;(\ref{reexpand}) for $L$ and (\ref{Bm}) for $L^{m+1}$ and $L^m$\,,
and collecting only positive contributions, we come to
\begin{equation}
\label{recurB}
B_{m+1} = \partial\circ B_m - \sum_{i=1}^m\sigma_{i}^{(1)}B_{m-i}
 - \sigma_{1}^{(m)}\,.
\end{equation}
Then, acting by both sides upon $\psi$\,, we get rid of $B$ and $\psi$
via (\ref{Baker}) and (\ref{dlnpsi})\,, and arrive at a general recursion
relation for $\sigma_i^{(m)}$\,:
\begin{equation}
\label{recurgen}
\sigma_i^{(m+1)} = \partial_m\sigma_i^{(1)} + \sigma_{i+1}^{(m)}
 + \sigma_{m+i}^{(1)} - \sum_{j=1}^{m-1}\sigma_j^{(1)}\sigma_i^{(m-j)}
 + \sum_{j=1}^{i-1}\sigma_j^{(1)}\sigma_{i-j}^{(m)}\,.
\end{equation}
Setting $i=n-m$ and summing up in $m$ cancels bilinear parts and
yields exactly (\ref{recurs1})\,.

An important corollary of (\ref{set2}) is that for any $k,m$
\begin{equation}
\label{}
\partial_k\partial_m\psi_1 = \partial_k\sigma_{1}^{(m)} = \partial(\ldots)\,,
\end{equation}
and we are prompted to make a definition
\begin{equation}
\label{tau}
\psi_1(t) \doteq -\partial\ln\tau(t)\,.
\end{equation}
This is the first appearance of the famous $\tau$-function. We see that
\begin{equation}
\label{}
\sigma_{1}^{(m)} = \partial_m\psi_1 = -\partial_m\partial \ln\tau\,, \ \ \ \ \
u_1 = - \sigma_{1}^{(1)} = \partial^2\ln\tau\,.
\end{equation}
Moreover, the recursion relation (\ref{recurs1}) can be explicitly resolved
for $\sigma_{n}^{(1)}$ in terms of $\tau$\,-function via the Schur
polynomials $p_n(t)$\,, which are defined (for natural $n$) as follows:
\begin{equation}
\label{Schur}
e^{\xi(t,z)}\doteq \sum_{n=0}^{\infty}z^n p_n(t)\,, \ \ \ \
\xi(t,z) = \sum_{i=1}^{\infty}z^i t_i\,.
\end{equation}
In particular, $p_0(t)=1\,, \ p_1(t) = t_1\,, \ p_2(t) =
\frac{1}{2}t_1^2 + t_2\,, \ p_3(t) = \frac{1}{6}t_1^3 + t_1 t_2 + t_3 $\,.
The following properties are readily deduced:
\begin{equation}
\label{propSchur}
\partial_m p_n(t) = p_{n-m}(t)\,, \ \ \ \
n p_n(\tilde{t}) = \sum_{m=1}^{n}t_m p_{n-m}(\tilde{t})\,,
\end{equation}
where tilde traditionally indicates a special scaling transformation of
(infinite number of) arguments of $p_n$\,,
\begin{equation}
\label{tilde}
\tilde{t} = \{t_1,\frac{t_2}{2},\frac{t_3}{3},\ldots\}\,, \ \ \ \
\tilde{\partial} = \{\partial_1,\frac{\partial_2}{2},
 \frac{\partial_3}{3},\ldots\}\,,
\end{equation}
so that
\begin{equation}
\label{tilSchur}
\exp\left(\sum_{n=1}^{\infty}\frac{t_n}{n} z^n\right)
= \sum_{n=0}^{\infty} p_n(\tilde{t}) z^n\,.
\end{equation}
The second equality in (\ref{propSchur}) follows from an observation
that multiplying (\ref{tilSchur}) by $1\!+\xi(t,z)$ is equivalent to
differentiating it by $z\frac{d}{dz}$\,.

The explicit solution of (\ref{recurs1}) mentioned above reads
\begin{equation}
\label{sigma1}
\sigma_{n}^{(1)} = p_n(-\tilde{\partial})\,\partial\ln\tau\,.
\end{equation}
To verify it, we replace $t_m$ in (\ref{propSchur}) by $-\partial_m$\,,
\begin{equation}
\label{recurpn}
n p_n(-\tilde{\partial})
= -\sum_{m=1}^{n-1}\partial_m p_{n-m}(-\tilde{\partial}) - \partial_n\,,
\end{equation}
apply this to $\partial\ln\tau$, and then compare with (\ref{recurs1})\,.

Now we see from $\partial\psi_i=\sigma_i^{(1)}$ that, up to irrelevant
$x$-independent terms,
\begin{equation}
\label{lnpsi2}
\ln\psi(t,z) = \sum_{n=1}^{\infty}z^n t_n
 + \sum_{i=1}^{\infty}z^{-i}p_i(-\tilde{\partial})\ln\tau
= \xi(t,z) + \ln\tau(\{t_m-\frac{1}{mz^m}\}) - \ln\tau(t)
\end{equation}
because
\begin{equation}
\label{}
f(t_1+\frac{1}{z}\,,t_2+\frac{1}{2z^2}\,,\ldots)
 = \exp\left(\sum_{n=1}^{\infty}\frac{\partial_n}{n z^n}\right)f(t)
 = \sum_{n=0}^{\infty} z^{-n}p_n(\tilde{\partial})f(t)\,.
\end{equation}
Thus, we are able to express the Baker function $\psi$ in terms of $\tau$\,:
\begin{equation}
\label{psi}
\psi(t,z) = e^{\xi(t,z)}\,\frac{\tau(\{t_m-\frac{1}{mz^m}\})}{\tau(t)}\,.
\end{equation}
Using again (\ref{lnpsi2}) and (\ref{recurpn}), we can also `invert' this
relation and express derivatives of $\tau$ in terms of $\psi$\,:
\begin{equation}
\label{}
\partial_n\ln\tau(t) = - \text{Res}_z\,z^n(\sum_{m=1}^{\infty}
z^{-m-1}\partial_m - \partial_z)\,\ln w(t,z)\,,
\end{equation}
where $\text{Res}_z$ means the coefficient of $z^{-1}$, \,and
\begin{equation}
\label{w}
\psi(t,z) = w(t,z)\,e^{\xi(t,z)}\,, \ \ \ \ \
w(t,z) = 1 + \sum_{i=1}^{\infty}w_i(t)z^{-i}\,.
\end{equation}
It should be noted here that a class of valid objects to act upon with
pseudodifferential operators according to the rules
$\partial^m e^{\xi(t,z)}=z^m e^{\xi(t,z)}$ and (\ref{diff}) is given by
$(\sum_{n\leqslant N}z^n b_n(t))\,e^{\xi(t,z)}$.

Due to (\ref{w}), it seems now natural to try the following representation
for $\psi$\,:
\begin{equation}
\label{psiW}
\psi(t,z) = W e^{\xi(t,z)}\,,
\end{equation}
where the \textit{dressing operator} $W$ is of the form
\begin{equation}
\label{W}
W = 1 + \sum_{i=1}^{\infty} w_i(t)\,\partial^{-i}\,.
\end{equation}
Owing to
$\partial_n e^{\xi(t,z)} = \partial^n e^{\xi(t,z)} = z^n e^{\xi(t,z)}$\,,
we see from (\ref{lnpsi}), (\ref{lnpsi2}), (\ref{psiW}) and (\ref{W}), that
$w_i$\,, $\psi_i$ and derivatives of $\tau$ are related by
\begin{equation}
\label{}
\sum_{i=1}^{\infty}z^{-i}\psi_i = \ln\,(1 + \sum_{i=1}^{\infty}z^{-i}w_i)
 = \sum_{i=1}^{\infty}z^{-i}p_i(-\tilde{\partial})\ln\tau\,.
\end{equation}

To guarantee (\ref{Baker}), operator $W$ should satisfy the following
``dressing" relations:
\begin{equation}
\label{dress}
L = W\partial\circ W^{-1}\,, \ \ \ \ \ \
\partial_n - B_n = W(\partial_n - \partial^n)\circ W^{-1}\,.
\end{equation}
The last of them reveals the nature of commutativity of the operators
$\partial_n - B_n$\,: they are nothing more than dressed counterparts
of the (trivially commuting) operators $\partial_n - \partial^n$\,.
Eqs. (\ref{dress}) can also be rewritten in the form involving $W$ only,
\begin{equation}
\label{eqW}
\partial_n W = -(W\partial^n\circ W^{-1})_-\,W\,.
\end{equation}

It looks quite feasible that an analogous equation for the
$\tau$\,-function in closed form also exists. This is really so:
one can obtain a fundamental \textit{bilinear identity}
\begin{equation}
\label{bilintau}
\text{Res}_z\,\tau(\{t_m-\frac{1}{mz^m}\})\,\tau(\{t'_m+\frac{1}{mz^m}\})
\,e^{\xi(t-t'\!,\,z)} = 0\,.
\end{equation}
To derive (\ref{bilintau}), it is useful to introduce a so-called adjoint
Baker function
\begin{equation}
\label{psi*}
\psi^*(t,z)\doteq (W^*)^{-1}e^{-\xi(t,z)}
 = w^*(t,z)\,e^{-\xi(t,z)}\,.
\end{equation}
In $W^*$\,, the asterisk (formal conjugation) changes sign of $\partial$
and the order of all operators,
\begin{equation}
\label{W*}
W^* = 1 + \sum_{i=1}^{\infty}(-\partial)^{-i}\circ w_i(t)\,, \ \ \ \
(W^*)^{-1} = 1 + \sum_{i=1}^{\infty} w^*_i(t)\,(-\partial)^{-i}\,,
\end{equation}
whereas $w^*$ is a mere notation. In analogy with the above reasoning,
one can derive
\begin{equation}
\label{w*}
w^*(t,z) = 1 + \sum_{i=1}^{\infty}w^*_i(t)\,z^{-i}
 = \frac{\tau(\{t_m+\frac{1}{mz^m}\})}{\tau(t)}\,,
\end{equation}
and thus rewrite the bilinear identity as follows:
\begin{equation}
\label{bilinw}
\text{Res}_z\,\psi(t,z)\,\psi^*(t',z) = 0\,.
\end{equation}
Bearing in mind the Taylor expansion in $t-t'$, it is evidently sufficient
to show that
\begin{equation}
\label{bilin1}
\text{Res}_z\,\partial^m\psi(t,z)\,\psi^*(t,z)=0\,, \ \ \ \ m\geqslant0\,,
\end{equation}
because other derivatives ($\partial_{n>1}$) can be treated
via (\ref{Baker}). First, for any two pseudodiffe\-rential operators
$P=\sum a_i \partial^i$ and $Q=\sum b_i \partial^i$ one proves by direct
calculation that
\begin{equation}
\label{Lemma1}
\text{Res}_z (Pe^{xz})(Qe^{-xz}) = \text{Res}PQ^*
 = \sum_{i+j=-1}(-)^j a_i b_j
\end{equation}
(recall that $\text{Res}=\text{Res}_\partial$)\,.
Now (\ref{bilin1}) follows:
\begin{multline}
\label{}
\text{Res}_z\,\partial^m\psi(t,z)\,\psi^*(t,z)
 = \text{Res}_z\,\partial^m(W e^{\xi(t,z)})\,(W^*)^{-1}e^{-\xi(t,z)} \\
 = \text{Res}_z\,(\partial^m\circ W e^{xz})\,(W^*)^{-1}e^{-xz}
= \text{Res}\,\partial^m\!\circ\!W W^{-1}  = \text{Res}\,\partial^m = 0\,.
\end{multline}

Conversely, (\ref{Baker}) can be derived from (\ref{bilinw}). In brief:
given $\psi$, we reconstruct $W$ by (\ref{W}), define $L$ by (\ref{dress}),
show that $L\psi=z\psi$ and $(\partial_n-B_n)\psi=(\partial_n-L^n+L_-^n)\psi
=e^{\xi}\mathcal{O}(z^{-1})$\,. But, due to
$\text{Res}_z(\partial_n-B_n)\psi(t,z)\cdot\psi^*(t',z)=0$\,,
this implies $(\partial_n-B_n)\psi=0$ (it is shown by setting $t'=t$
after differentiating by $t' \ \ 0,1,\ldots$ times).
Thus, bilinear identities (\ref{bilintau}), (\ref{bilinw}) turn out to be
equivalent to entire KP hierarchy (\ref{main}).
We shall see below that bilinear relations
of this kind are naturally interpreted within the fermionic
representation of the $\tau$ and Baker functions.

\vspace{.3cm}

\begin{center}
\large\textbf{Fermionic Fock space}
\end{center}

\vspace{.1cm}

Remarkably, central objects of the KP hierarchy considered above
(Baker's and $\tau$-functions, bilinear equations, and some others)
can be represented, and efficiently dealt with, in the framework
of free fermionic picture, via so-called Fermi-Bose correspondence.
This also reveals the underlying group theory structure of integrable
hierarchies.

We begin with description of an appropriate fermionic Fock space.
Consider an algebra of the free fermion operators,
\begin{equation}
\label{free}
[\psi_m,\psi_n]_+ = [\bar{\psi}_m,\bar{\psi}_n]_+ = 0\,,
\ \ \ \ \ \ [\psi_m,\bar{\psi}_n]_+ =\delta_{mn}\,,
\end{equation}
with arbitrary integer indices $m,n$\,. We shall call
$\psi_n \ (\equiv \psi_n^+)$ the wedging (or creation) and
$\bar{\psi}_n \ (\equiv \psi_n^-)$ the contraction (or annihilation)
operators. Of course, these are generators of the Clifford algebra:
\begin{equation}
\label{Cliff}
e_n = \psi_n + \bar{\psi}_n\,, \ \ \
e_{\bar n} = i(\psi_n - \bar{\psi}_n)\,, \ \ \ \
[e_{\alpha},e_{\beta}]_+ = 2\delta_{\alpha\beta}\,.
\end{equation}
Let $|\,\Omega\rangle$ (fake vacuum) be a state annihilated by each $\bar{\psi}_n$\,,
and $|\,0\rangle$ (physical vacuum) the following formal object:
\begin{equation}
\label{vac0}
|\,0\rangle = \psi_{-1}\psi_{-2}\ldots|\,\Omega\rangle\,, \ \ \ \ \ \
\bar{\psi}_n |\,\Omega\rangle = 0\,.
\end{equation}
Then
\begin{equation}
\label{creann}
\psi_{n<0}|\,0\rangle = \bar{\psi}_{n\geqslant0}|\,0\rangle = 0\,,
\end{equation}
and we can interpret the free fermionic operators as follows:

$\psi_{n\geqslant0}$ \ \ \ \ \ -- \ creation of a particle

$\bar{\psi}_{n\geqslant0}$ \ \ \ \ \ -- \ annihilation of a particle

$\bar{\psi}_{n<0}$ \ \ \ \ \ -- \ creation of a hole

$\psi_{n<0}$ \ \ \ \ \ -- \ annihilation of a hole

\noindent
It is also useful to introduce, for any integer $m$, the vectors
\begin{equation}
\label{mvac0}
|\,m\rangle = \psi_{m-1}\psi_{m-2}\ldots|\,\Omega\rangle\,,
\end{equation}
so that
\begin{equation}
\label{}
\psi_{n<m}|\,m\rangle = \bar{\psi}_{n\geqslant m}|\,m\rangle = 0\,.
\end{equation}

The fermionic Fock space $\mathsf F$ is spanned by vectors $|\,u\rangle$
which are obtained by application of a finite number of $\psi_m,\bar{\psi}_n$
operators to $|\,0\rangle$\,. So, $|\,\Omega\rangle$ is not in $\mathsf F$\,.
We also introduce the notion of \textit{charge} to characterize the total
number of operators $\psi$ used (in comparison with the vacuum
$|\,0\rangle$)\,. Thus, the charges of $\psi_n,\bar{\psi}_n$ and $|\,m\rangle$
are $+1, -1$ and $m$\,, respectively.

Let us now define the fermion fields as the following formal expansions:
\begin{gather}
\label{psi0}
\psi(z) = \sum_{n}z^n\psi_n = \psi_-(z) + \psi_+(z)\,, \ \ \
\psi_-(z) = \sum_{n<0}z^n\psi_n\,, \ \ \
\psi_+(z) = \sum_{n\geqslant0}z^n\psi_n\,, \\
\notag
\bar{\psi}(z) = \sum_{n}z^{-n-1}\bar{\psi}_n
 = \bar{\psi}_+(z) + \bar{\psi}_-(z)\,, \ \ \
\bar{\psi}_+(z) = \sum_{n<0}z^{-n-1}\bar{\psi}_n\,, \ \ \
\bar{\psi}_-(z) = \sum_{n\geqslant0}z^{-n-1}\bar{\psi}_n\,, \\
\label{delta0}
[\psi(z)\,,\,\bar{\psi}(w)]_+ = \sum_n w^n z^{-n-1} = \delta(z-w)\,,
\end{gather}
see (\ref{delta}) for a discussion of this definition of $\delta$-function.
We see that the $(-)$ part of a field corresponds to negative,
whereas the $(+)$ part to non-negative powers of $z$\,, but a more deep
mnemonic rule is that $+$ corresponds to creation and $-$ to annihilation:
\begin{equation}
\label{}
\psi(z)|\,0\rangle = \psi_+(z)|\,0\rangle\,, \ \ \ \ \
\bar{\psi}(z)|\,0\rangle = \bar{\psi}_+(z)|\,0\rangle\,.
\end{equation}

The duality (bra vs ket) properties of $\mathsf F$ are self-evident,
if we accept $\psi_n$ and $\bar{\psi}_n$ to be mutually Hermitean
conjugate. We denote $\langle 0\,|\ldots|\,0\rangle$ by
$\langle \ldots \rangle$ and use
$\langle m\,|\,m\rangle=1$\,. Obviously,
\begin{equation}
\label{}
\langle \bar{\psi}_m\psi_n\rangle = -\langle \psi_n\bar{\psi}_m\rangle
 + \delta_{mn} = \delta_{mn}\theta(n)\,, \ \ \ \ \ \ \theta(n) =
\begin{cases}
1 & \qquad (n\geqslant 0) \\ 0 & \qquad (n<0)
\end{cases}
\end{equation}
Another useful example of a matrix element is (see (\ref{expanzw}))
\begin{equation}
\label{}
\langle \psi(z)\bar{\psi}(w)\rangle =
\langle \bar{\psi}(z)\psi(w)\rangle
= \sum_{n\geqslant0}\,w^n z^{-n-1} = (z-w)_{w}^{-1}\,.
\end{equation}

The normal product is defined as usual: annihilation (with respect to the
physical vacuum $|\,0\rangle$) operators should not stand to the left of
the creation ones.
For $\varphi$ and $\chi$ linear in $\psi,\bar{\psi}$\,, it is
\begin{equation}
\label{normal}
:\varphi\chi: \ \doteq \ \varphi_+\chi - \chi\varphi_-
= \varphi\chi - \langle\varphi\chi\rangle
= \varphi\chi - \langle\varphi_-\chi_+\rangle\,.
\end{equation}
In particular,
\begin{equation}
\label{}
:\bar{\psi}_m\psi_n: \ =\  -:\psi_n\bar{\psi}_m: \
= \,\bar{\psi}_m\psi_n - \delta_{mn}\theta(n)\,.
\end{equation}

Now we introduce a bilinear current
\begin{equation}
\label{current}
J(z)\,= \ :\psi(z)\bar{\psi}(z): \ = \sum_{n}z^{-n-1}J_n\,,
\ \ \ \ \ J_n = \sum_{m}:\psi_m\bar{\psi}_{n+m}:
\end{equation}
In fact, the normal ordering is effective only for $J_0$\,:
\begin{equation}
\label{J0}
J_0 = \sum_{m\geqslant0}\psi_m\bar{\psi}_m
 - \sum_{m<0}\bar{\psi}_m\psi_m\,.
\end{equation}
The following holds:
\begin{equation}
\label{propJ}
(J_n)^+ = J_{-n}\,, \ \ \ \ J_0|\,m\rangle = m|\,m\rangle\,, \ \ \ \
J_{n>0}|\,m\rangle = 0\,,
\end{equation}
so that $J_0$ measures the charge. The last equality in (\ref{propJ})
is not surprising because $J_{n>0}$ effectively shifts
particles by $n$ positions down, which is clearly impossible in a `filled'
state $|\,m\rangle$\,. Moreover, for any Fock vector $|\,u\rangle$
and sufficiently large $n$\,,
\begin{equation}
\label{}
J_n|\,u\rangle = 0 \ \ \ \ \ \ \ (n\gg0)\,.
\end{equation}
Further,
\begin{gather}
[J_n, \psi_m] = \psi_{m-n}\,, \ \ \ \ \
[J_n, \bar{\psi}_m] = -\bar{\psi}_{m+n}\,, \\
\label{Jnpsi0}
[J_n, \psi(z)] = z^n\psi(z)\,, \ \ \ \ \
[J_n, \bar{\psi}(z)] = -z^n\bar{\psi}(z)\,.
\end{gather}
At last, the commutator of the current components has a typical
bosonic form:
\begin{multline}
\label{commJ}
[J_m,J_n] = [J_m,\,\sum_{k\geqslant0}\psi_k\bar{\psi}_{n+k}
 - \sum_{k<0}\bar{\psi}_{n+k}\psi_k] \\
= \sum_{k\geqslant0}(\psi_{k-m}\bar{\psi}_{n+k} - \psi_k\bar{\psi}_{n+m+k})
- \sum_{k<0}(-\bar{\psi}_{n+m+k}\psi_k + \bar{\psi}_{n+k}\psi_{k-m}) \\
= \sum_{l=-m}^{-1}(\psi_l\bar{\psi}_{l+n+m} + \bar{\psi}_{l+n+m}\psi_l)
= m\delta_{m,-n}\,.
\end{multline}

The current $J(z)$ contains infinite summation. To work with such operators
in the Fock space, we need some special techniques. By definition, any Fock
vector is \textit{finite} in the following sense: it differs from
$|\,0\rangle$ in a finite number of positions (filled or empty), or,
in other words, the maximum $|n|$ of a position whose status differs from
that in $|\,0\rangle$\,, is finite. When treating Fock vectors as infinite
columns (the occupation number representation), it is clear that a finite
vector has only finite number of nonzero components. We are interested in
the operators which send finite vectors to finite. These are described by
so-called finite matrices, which obey $A_{ij}=0$ for $|i-j|\gg 0$ (so
a finite matrix may actually have infinite number of nonzero elements).
Multiplication (and commutation) operations respect finiteness.

Now consider a class of operators in $\mathsf F$ of the form
\begin{equation}
\label{XA}
X_A = \sum_{mn}A_{mn}:\psi_m\bar{\psi}_n:
\end{equation}
with $A$ finite. Such operators carry zero charge, and send Fock vectors to
Fock vectors: normal ordering cuts off the regions $n\gg0$ and $m\ll0$\,,
and finiteness of $A$ completes the job. For instance,
\begin{equation}
\label{JviaX}
J_n = X_{A_n}\,, \ \ \ \ \ \ (A_n)_{mk} = \delta_{k,n+m}\,.
\end{equation}
One easily deduces
\begin{equation}
\label{Xpsi}
[X_A,\psi_n] = A_{mn}\psi_m\,, \ \ \ \ \ \
[X_A,\bar{\psi}_n] = -A_{nm}\bar{\psi}_m\,,
\end{equation}
and
\begin{equation}
\label{ext}
[X_A,X_B] = X_{[A,B]} + \omega(A,B)
\end{equation}
where the 2-cocycle
\begin{equation}
\label{omega}
\omega(A,B) = -\omega(B,A)
= \sum_{ij}A_{ij}B_{ji}(\theta(j)-\theta(i))
\end{equation}
is well defined for finite $A$ and $B$\,. For example,
\begin{equation}
\label{}
\omega(A_m,A_n) = m\delta_{m,-n}\,.
\end{equation}
Thus we come to a centrally extended Lie algebra
\begin{equation}
\label{glinfty}
\text{gl}_\infty = \{X_A\} \oplus \mathbb{C}\,.
\end{equation}
Really, the basic commutator
\begin{equation}
\label{}
[\psi_m\bar{\psi}_n,\psi_p\bar{\psi}_q]
= \delta_{pn}\psi_m\bar{\psi}_q - \delta_{qm}\psi_p\bar{\psi}_n
\end{equation}
copies the commutator of the (gl)-generators
$(E_{mn})_{ij}=\delta_{im}\delta_{jn}$\,.

The corresponding Lie group $\mathbf{G}$ (a subgroup of the Clifford group)
consists of invertible elements $g$ of the Clifford algebra obeying
\begin{equation}
\label{G}
gVg^{-1}\subset V\,, \ \ \ \ \ \ g\bar{V}g^{-1}\subset \bar{V}\,,
\end{equation}
with $V$ being a linear space spanned by $\psi_n$ and $\bar{V}$ by
$\bar{\psi}_n$\,. Its connected subgroup containing unity is formed by
the elements of the type \,$\exp(X_A)$\,.

Consider as an example the Clifford algebra generated by $[a,a^\dag]_+=1,
 \ \,aa=a^\dag a^\dag=0$\,. Its basis is $\{1,a,a^\dag,a^\dag a\}$\,,
zero-charge elements are 1 and $a^\dag a$\,, whereas $V$
and $\bar{V}$ are one-dimensional spaces along $a$ and $a^\dag$,
respectively. The elements of the group of our interest are
\begin{equation}
\label{example}
\alpha(1 + \beta a^\dag a) \ \ \ \ \ \text{with} \ \ \
 \alpha \neq 0\,, \ \beta\neq -1\,.
\end{equation}
Namely,
\begin{equation}
\label{}
(1 + \beta a^\dag a)^{-1} = (1-\frac{\beta}{1+\beta}a^\dag a)\,.
\end{equation}
For $\beta>-1$ elements (\ref{example}) admit exponential form due to
\begin{equation}
\label{expform}
e^{\gamma a^\dag a} = 1 + (e^\gamma -1)a^\dag a\,,
\end{equation}
otherwise they don't, being in the second connected component
(without unity).

We claim that any element of the orbit $\mathbf G|\,0\rangle$ obeys
bilinear relations specific to the $\tau$-function of the KP hierarchy,
and thus is nothing more than a fermionic appearance of this function.
To prove this (and other) statements about the properties of $\tau$
and Baker functions, and to relate their fermionic representations
with more conventional ones, we need to discuss the Fermi-Bose correspondence.

\vspace{.3cm}

\begin{center}
\large\textbf{Fermi-Bose correspondence}
\end{center}

\vspace{.1cm}

There is an isomorphism between the fermionic Fock space $\mathsf F$
and the (bosonic Fock) space $\mathsf B$ of polynomials in
$t_n,y,y^{-1}$\,. Let us define
\begin{equation}
\label{H(t)}
H(t) = \sum_{n>0}t_n J_n\,.
\end{equation}
It is easily seen that
\begin{equation}
\label{propH}
H(t)|\,m\rangle = 0\,, \ \ \ \ \ [H(t),J_n] = (\theta(n)-1)n t_{-n}\,.
\end{equation}
Further useful properties of $H(t)$ are formulated in terms of $\xi(t,z)$
(\ref{Schur})\,:
\begin{equation}
\label{}
[H(t),\psi(z)] = \xi(t,z)\psi(z)\,, \ \ \ \ \
[H(t),\bar{\psi}(z)] = -\xi(t,z)\bar{\psi}(z)\,,
\end{equation}
and, consequently,
\begin{equation}
\label{commH}
e^{H(t)}\psi(z)e^{-H(t)} = e^{\xi(t,z)}\psi(z)\,, \ \ \ \ \
e^{H(t)}\bar{\psi}(z)e^{-H(t)} = e^{-\xi(t,z)}\bar{\psi}(z)\,.
\end{equation}
With the use of Schur polynomials (\ref{Schur}) one can also write
\begin{gather}
\label{commH1}
e^{H(t)}\psi_n e^{-H(t)} = \sum_{m=0}^{\infty}p_m(t)\psi_{n-m}\,, \\
\label{commH2}
e^{H(t)}\bar{\psi}_n e^{-H(t)} = \sum_{m=0}^{\infty}p_m(-t)\bar{\psi}_{n+m}\,.
\end{gather}

Now, the \textit{bosonization map} is
\begin{equation}
\label{bos}
B(|\,u\rangle)\doteq \sum_{m}y^m \langle m|e^{H(t)}|\,u\rangle\,.
\end{equation}
For any Fock vector $|\,u\rangle$\,, $B(|\,u\rangle)$ is a
polynomial in $t_n,y,y^{-1}$\,, which is identical
zero only for $|\,u\rangle=0$\,. Clearly, $B(|\,m\rangle)=y^m$\,.
Also, if $|\,u\rangle$ carries a definite charge $m$\,, only
the $y^m$-\,term of the sum in (\ref{bos}) survives. Furthermore, if
$|\,u\rangle$ is a monomial in $\psi,\bar{\psi}$ acting upon
$|\,0\rangle$\,, then its Bose-counterpart can be explicitly written
in terms of the Schur polynomials. Really, such a monomial
$|\,\varphi\rangle$ is fully characterized by its charge (say, $m$)
and a \textit{partition} $\lambda_\varphi$ which is a finite
non-increasing sequence of natural numbers $\{\lambda_1,\lambda_2,\ldots\}$\,,
indicating the differences in positions of particles in
$|\,\varphi\rangle$ and $|\,m\rangle$\,, beginning from the top.
A general formula
\begin{equation}
\label{bosmonom}
B(|\,\varphi\rangle) = y^m \langle m|e^{H(t)}|\,\varphi\rangle
 = y^m p_{\lambda_\varphi}(t)
\end{equation}
results from repeated use of (\ref{commH1}).
The Schur polynomials $p_\lambda(t)$ indexed by
partitions are defined through elementary Schur polynomials
(\ref{Schur}) as follows (we assume $p_{n<0}(t)=0$):
\begin{equation}
\label{plambda}
p_\lambda(t) = \det (p_{\lambda_i+j-i}(t))\,, \ \ \ \
 1\leqslant i,j\leqslant\sum_k\lambda_k\,.
\end{equation}
For example, the state $|\,\varphi\rangle=\psi_m\psi_{m-1}|\,m\!-\!2\rangle$
has $\lambda_\varphi=\{1,1\}$ and
\begin{equation}
\label{}
p_{\lambda_\varphi}(t) = \det\left(
 \begin{array}{cc} p_1(t) & p_2(t) \\ p_0(t) & p_1(t) \end{array}\right)
 = p_1^2(t) - p_0(t) p_2(t)\,.
\end{equation}

Let us now derive, for later use, a formula
\begin{equation}
\label{key}
\langle m\!+\!1|\psi(z)|\,\varphi\rangle
 = z^m p_{\lambda_\varphi}(\{-\frac{1}{nz^n}\})
 = z^m \langle m|e^{-H(\{\frac{1}{nz^n}\})}|\,\varphi\rangle\,.
\end{equation}
To produce a nonzero contribution, a monomial $|\,\varphi\rangle$
(of charge $m$) should be equal to $|\,m\rangle$\,, or
display exactly one vacancy (hole) in its
`body' to be filled in by some $\psi_n$ from $\psi(z)$\,. Namely
($N=0,1,2,\ldots$)
\begin{equation}
\label{111}
|\,\varphi\rangle = \psi_m\psi_{m-1}\ldots\psi_{m-N+1}|\,m\!-\!N\rangle \ \
 \ \ \Rightarrow \ \ \ \
\langle m\!+\!1|\psi(z)|\,\varphi\rangle = (-)^N z^{m-N}\,.
\end{equation}
In other words, a partition $\lambda_\varphi$ may contain only 1's. To
compute the corresponding determinant in (\ref{plambda}), observe that
\begin{equation}
\label{}
\sum_{n=0}^{\infty}p_n(\{-\frac{1}{mz^m}\})\,w^n
 = \exp\,(-\sum_{n=1}^{\infty}\frac{w^n}{nz^n})
 = \exp\ln(1-\frac{w}{z}) = 1-\frac{w}{z}\,,
\end{equation}
so
\begin{equation}
\label{vanish}
p_1(\{-\frac{1}{mz^m}\}) = -\frac{1}{z}\,, \ \ \ \
p_n(\{-\frac{1}{mz^m}\}) = 0 \ \ \ \ \text{for} \ \ n\geqslant2\,.
\end{equation}
Now the determinants in $p_{\lambda_\varphi}$ in eq. (\ref{key}) are
easily evaluated. They either agree with (\ref{111}), for appropriate
$|\,\varphi\rangle$, or vanish due to (\ref{vanish}), in both cases
verifying (\ref{key}). Due to $|\,\varphi\rangle$ being arbitrary
monomial we, in fact, have proved the following useful formula:
\begin{equation}
\label{key2}
\langle m|\psi(z) = z^{m-1}\,\langle m\!-\!1|e^{-H(\{\frac{1}{nz^n}\})}\,.
\end{equation}
Analogously, one derives
\begin{equation}
\label{key3}
\langle m|\bar{\psi}(z)
 = z^{-m-1}\,\langle m\!+\!1|e^{H(\{\frac{1}{nz^n}\})}\,.
\end{equation}

Let us now proceed with bosonization of various fermionic operators.
From (\ref{propH}), due to $\langle m|J_0=m\langle m|\,, \
\langle m|J_{n<0}=0$\,, we obtain
\begin{equation}
\label{BJ0}
B(J_0|\,u\rangle) = y\partial_y B(|\,u\rangle)
\end{equation}
(hence, $q^{J_0}$ bosonizes to the operation $y\rightarrow qy$, \
i.e., $B(q^{J_0}|\,u\rangle)(t,y) = B(|\,u\rangle)(t,qy)$), and
\begin{equation}
\label{BJn}
B(J_n|\,u\rangle) = \partial_n B(|\,u\rangle)\,, \ \ \
B(J_{-n}|\,u\rangle) = n t_n B(|\,u\rangle) \ \ \ \ (n>0)\,.
\end{equation}
One concludes that components of the current $J(z)$ are represented in
$\mathsf B$ as follows $(n>0)$\,:
\begin{equation}
\label{bosJ}
J_0 \sim y\partial_y = \partial_{\ln y}\,, \ \ \ \
J_n \sim \partial_n\,, \ \ \ \ J_{-n} \sim n t_n\,,
\end{equation}
which completely agrees with their commutation rules (\ref{commJ}).

It appears useful to introduce in $\mathsf F$ an operator $Y$
which augments charge by one and bosonizes to multiplication by $y$: \
$B(Y|\,u\rangle) = yB(|\,u\rangle)$\,. This is achieved via
the definition
\begin{equation}
\label{K}
Y\,\psi_n = \psi_{n+1}\,Y\,, \ \ \ \ Y^+ = Y^{-1}
\end{equation}
that leads to the following relations:
\begin{gather}
Y\bar{\psi}_n = \bar{\psi}_{n+1} Y\,, \ \ \ \ \
Y|\,m\rangle = |\,m+1\rangle\,, \ \ \ \ \
\langle m|\,Y = \langle m-1|\,, \\
\label{JK}
[J_n,Y] = 0 \ \ \ \ (n\neq0)\,, \ \ \ \ \ \ \
[J_0,Y] = Y\,, \ \ \ \ \ \ \
q^{J_0}Y = q\,Y\,q^{J_0}\,.
\end{gather}
Taken in the form $[J_0\,,\ln Y]\sim[\partial_{\ln y}\,,\ln y]=1$\,, this
suggests to interpret $\ln Y$ as a creation and $J_0$ as an annihilation
operator.

Now we are in a position to bosonize fermion fields.
\begin{multline}
\label{bosd}
B(\psi(z)|\,u\rangle) = \sum_m y^m \langle m|e^{H(t)}\psi(z)|\,u\rangle
 = e^{\xi(t,\,z)}\sum_m y^m \langle m|\psi(z)e^{H(t)}|\,u\rangle \\
= e^{\xi(t,\,z)}\sum_m y^m z^{m-1}
  \langle m\!-\!1|e^{H(\{t_n-\frac{1}{nz^n}\})}|\,u\rangle
 = e^{\xi(t,\,z)}e^{-\xi(\tilde{\partial},\,z^{-1})}
  \sum_m y^m z^{m-1}\langle m\!-\!1|e^{H(t)}|\,u\rangle \\
= e^{\xi(t,\,z)}e^{-\xi(\tilde{\partial},\,z^{-1})}y
  \sum_m (zy)^m \langle m|e^{H(t)}|\,u\rangle
 = e^{\xi(t,\,z)}e^{-\xi(\tilde{\partial},\,z^{-1})}\,y\,
  z^{y\partial_y}B(|\,u\rangle)\,.
\end{multline}
Performing a similar calculation for $\bar\psi$ we conclude that
\begin{equation}
\label{bospsi}
\psi^{\pm}(z) \sim e^{\pm\xi(t,\,z)}e^{\mp\xi(\tilde{\partial},\,z^{-1})}
\,y^{\pm1}\,z^{\pm y\partial_y} =
e^{\pm\sum_{n>0}z^n t_n}\,e^{\mp\sum_{n>0}\frac{\partial_n}{n z^n}}
\,y^{\pm1}\,z^{\pm y\partial_y}\,.
\end{equation}
In view of the bosonized form of $J_n$\,, this implies
\begin{equation}
\label{psiF}
\psi^{\pm}(z) = \exp(\pm\sum_{m=1}^{\infty}\frac{z^m}{m}J_{-m})\,
 \exp(\mp\!\sum_{n=1}^{\infty}\frac{J_n}{nz^n})\ Y^{\pm1}\,z^{\pm J_0}
\end{equation}
and may be considered as properly ordered bosonic
counterpart of \,$:\!e^{\pm\varphi(z)}\!:$ with
\begin{equation}
\label{phi0}
\varphi(z) = \sum_{n\neq0}\frac{J_n}{-n}z^{-n}
 + J_0 \ln z + \ln Y\,, \ \ \ \ \ \frac{d}{dz}\varphi(z) = J(z)\,.
\end{equation}

Clearly, in our derivation (\ref{bosd}) of the bosonization formulas
(\ref{bospsi}),(\ref{psiF}) for the fermion fields, the relations
(\ref{key2}),(\ref{key3}) play the crucial role. It should be noted that
the inverted calculation, i.e., deducing (\ref{key2}),(\ref{key3}) from
(\ref{psiF}), is even simpler. So, in principle, all the construction could
be started with deriving (\ref{psiF}) from the `first principles':
namely, from the commutators (\ref{Jnpsi0}). For example, the fermionic
counterpart of $e^{-\xi(t,z)}\psi(z)\,e^{\xi(\tilde{\partial},\,z^{-1})}$
proves to commute with $J_n$ for $n\neq 0$ and is thus equal to $Yf(z,J_0)$
(because in $\mathsf B$ this means commutativity with $t_n$ and
$\partial_n$\,, and, as a result, independence of both)\,. A function $f$
is found (say, using $\langle m+1|\psi(z)|\,m\rangle=z^m$) to be $z^{J_0}$,
which justifies (\ref{psiF}).

Exponential operators like those encountered in
(\ref{bospsi}) and (\ref{psiF})\,,
\begin{equation}
\label{vert}
V^\alpha(z) = \exp(\alpha\!\sum_{n>0}z^n t_n)
\,\exp(-\alpha\!\sum_{n>0}\frac{\partial_n}{n z^n})
\,y^{\alpha}\,z^{\alpha y\partial_y}
\end{equation}
as well as their $\mathsf F$-counterparts, are known as
\textit{vertex operators}. Of course, $\psi^{\pm}(z)\sim V^{\pm1}(z)$\,.
From (\ref{JK}) and the well-known identity
\begin{equation}
\label{abba}
[a,b] = \lambda\in\mathbb{C} \ \ \ \ \Longrightarrow \ \ \ \
e^a e^b = e^\lambda e^b e^a
\end{equation}
we find \ $z^{\alpha y\partial_y}\,y^\beta = z^{\alpha\beta}\,y^\beta\,
z^{\alpha y\partial_y}$\,,
\begin{gather*}
\left[-\alpha\!\sum_{m=1}^{\infty}\frac{\partial_m}{mz^m}\,,\,
  \beta\!\sum_{n=1}^{\infty}w^n t_n\right]
 = -\alpha\beta\!\sum_{n=1}^{\infty}\frac{w^n}{nz^n}
 = \alpha\beta\ln(1-\frac{w}{z})\,, \\
\exp(-\alpha\!\sum_{m=1}^{\infty}\frac{\partial_m}{mz^m})\,
\exp(\beta\!\sum_{n=1}^{\infty}w^n t_n)
 =\frac{(z-w)_w^{\alpha\beta}}{z^{\alpha\beta}}
\exp(\beta\!\sum_{n=1}^{\infty}w^n t_n)\,
\exp(-\alpha\!\sum_{m=1}^{\infty}\frac{\partial_m}{mz^m})\,,
\end{gather*}
and finally obtain
\begin{multline}
\label{VV}
V^\alpha(z)\,V^\beta(w)
 = (z-w)_w^{\alpha\beta}\,:\!V^\alpha(z)\,V^\beta(w)\!: \\
 = (z-w)_w^{\alpha\beta}\,e^{\alpha\xi(t,\,z)+\beta\xi(t,\,w)}\,
e^{-\alpha\xi(\tilde{\partial},\,z^{-1})-\beta\xi(\tilde{\partial},\,w^{-1})}
\,y^{\alpha+\beta}(z^\alpha w^\beta)^{y\partial_y}\,.
\end{multline}

Let us consider some particular cases.
Using (\ref{VV}) for $\alpha=1, \,\beta=-1$ together with an identity
$(z-w)_w^{-1}-(z-w)_z^{-1}=\delta(z-w)$ (see (\ref{grenze})\,)\,,
one easily checks (\ref{delta0})\,:
\begin{equation}
\notag
[\psi(z)\,,\,\bar{\psi}(w)]_+ \,\sim\, e^{\xi(t,\,z)-\xi(t,\,w)}
e^{-\xi(\tilde{\partial},\,z^{-1})+\xi(\tilde{\partial},\,w^{-1})}
\,(\frac{z}{w})^{y\partial_y}\,((z-w)_w^{-1}+(w-z)_z^{-1})
 = \delta(z-w)\,.
\end{equation}
Analogously, one can verify that, for example,
\begin{equation}
\label{}
\psi(z)\,\psi(w)\sim (z-w)\,e^{\xi(t,\,z)+\xi(t,\,w)}
e^{-\xi(\tilde{\partial},\,z^{-1})-\xi(\tilde{\partial},\,w^{-1})}
y^2\,(zw)^{y\partial_y}
\end{equation}
is manifestly antisymmetric, as it should be due to fermionic nature
of $\psi$\,.

\vspace{.3cm}

\begin{center}
\large\textbf{KP hierarchy via free fermions}
\end{center}

\vspace{.1cm}

Now we are to justify our claim that any $g\in\mathbf G$ is the
fermionic counterpart of a $\tau$-function, i.e., of
some solution of the (bilinear identities of) KP hierarchy. Namely,
we offer the following bosonized expressions as candidates for
the role of $\tau$ and Baker functions:
\begin{gather}
\label{tauB}
\tau(t) \equiv \tau(t;g) = \langle 0|e^{H(t)}g|\,0\rangle\,, \\
\label{psiB}
\psi(t,z) =
\frac{\langle1|e^{H(t)}\psi(z)g|\,0\rangle}{\tau(t;g)}\,, \\
\label{psi*B}
\psi^*(t,z) =
\frac{\langle -\!1|e^{H(t)}\bar{\psi}(z)g|\,0\rangle}{\tau(t;g)}\,.
\end{gather}
To begin with, we check the relation (\ref{psi}) using (\ref{commH})
and (\ref{key2})\,:
\begin{multline}
\label{}
\langle1|e^{H(t)}\psi(z)g|\,0\rangle
 = e^{\xi(t,z)}\langle1|\psi(z)e^{H(t)}g|\,0\rangle
 = e^{\xi(t,z)}\langle0|e^{H(t)-H(\{\frac{1}{nz^n}\})}g|\,0\rangle \\
= e^{\xi(t,z)}\langle0|e^{H(\{t_n-\frac{1}{nz^n}\})}g|\,0\rangle
 = e^{\xi(t,z)}\,\tau(\{t_n-\frac{1}{nz^n}\})\,.
\end{multline}
Similarly, using (\ref{key3}) serves to confirm (\ref{w*})\,.

Since the proper relations between the $\tau$ and Baker functions
are thus verified, it suffices to check the bilinear identity in the
form (\ref{bilinw})\,:
\begin{multline}
\label{}
\text{Res}_z\,\langle1|e^{H(t)}\psi(z)g|\,0\rangle\,
 \langle -\!1|e^{H(t')}\bar{\psi}(z)g|\,0\rangle
 = \sum_n\,\langle1|e^{H(t)}\psi_n g|\,0\rangle\,
 \langle -\!1|e^{H(t')}\bar{\psi}_n g|\,0\rangle \\
= \left(\langle1|e^{H(t)}\otimes\langle -\!1|e^{H(t')}\right)
\sum_n\,\psi_n g|\,0\rangle\otimes\bar{\psi}_n g|\,0\rangle\,.
\end{multline}
But any element $|\,u\rangle = g|\,0\rangle$ of the orbit
$\mathbf G|\,0\rangle$ obeys a bilinear relation
\begin{equation}
\label{bilinF}
\text{Res}_z\,\psi(z)|\,u\rangle\otimes\bar{\psi}(z)|\,u\rangle
 = \sum_{n}\psi_n|\,u\rangle\otimes\bar{\psi}_n|\,u\rangle = 0
\end{equation}
which is readily shown in the infinitesimal form $g=e^{X_A}\approx 1+X_A$
with the help of (\ref{Xpsi})\,:
\begin{multline}
\label{}
\sum_{n}\psi_n g|\,0\rangle\otimes\bar{\psi}_n g|\,0\rangle
\approx \sum_n\,(\psi_n|\,0\rangle\otimes\bar{\psi}_n|\,0\rangle
 + \psi_n X_A|\,0\rangle\otimes\bar{\psi}_n|\,0\rangle
 + \psi_n|\,0\rangle\otimes\bar{\psi}_n X_A|\,0\rangle) \\
= (\text{id}\otimes\text{id} + X_A\otimes\text{id} + \text{id}\otimes X_A)
 \,\sum_{n}\psi_n |\,0\rangle\otimes\bar{\psi}_n|\,0\rangle = 0
\end{multline}
because for any $n$ one of $\psi_n,\bar{\psi}_n$ must be an annihilation
operator. The bilinear identity (\ref{bilinw}) is thus proved. Its another
version (\ref{bilintau}), in terms of the $\tau$\,-function, follows from
(\ref{bilinF}) through the bosonization of fermion fields.

Of course, the orbit $\mathbf G|\,0\rangle$ provides a great supply of
solutions of bilinear identities (hence, of the KP hierarchy). The simplest
one is $\tau=1$ which corresponds to the vacuum state itself. Further, any
zero-charged monomial $|\,\varphi\rangle$ (see (\ref{bosmonom}))
lies in (a completion~of) the orbit,
therefore all Schur polynomials $p_{\lambda_\varphi}(t)$ are solutions.
Another important class of ($N$-\,soliton) solutions is produced by the (repeated)
action of exponents of vertex operators, related to quadratic
combinations of the Fermi fields, upon 1.

\newpage

\begin{center}
\large\textbf{II. \ VERTEX ALGEBRAS}
\end{center}

\vspace{.0cm}

\begin{center}
\large\textbf{Formal distributions and locality}
\end{center}

\vspace{.1cm}

Vertex algebras encode mathematical content of 2-dimensional
conformal quantum field theory (CFT${}_2$). Roughly speaking, chiral
quantum fields depending on a single complex variable $z$ are replaced
by operator-valued formal power series in an abstract variable $z$\,.
We begin with a brief exposition of the corresponding formalism.

Formal power series of the following type
($m,\,n\in\mathbb{Z}\,, \ a_n,\,a_{m,n}\in A$),
\begin{equation}
\label{distrib}
a(z) = \sum_{n}\frac{a_n}{z^{n+1}}\,, \ \ \ \
a(z,\,w) = \sum_{m,\,n}\frac{a_{m,n}}{w^{m+1}\,z^{n+1}}\,,
\end{equation}
are called $A$-valued formal distributions in one, two, or more
indeterminates $z,\,w,\,...$\,.
We can freely multiply formal distributions by polynomials,
but a product of two distributions in the same variable(s)
is not generally defined. Also, there is a problem in treating
negative powers of $(z-w)$ etc. as formal distributions.
To do it unambiguously, we will use (when needed) the notation like
\begin{equation}
\label{expanzw}
(z-w)_{w}^{n} \doteq \sum_{k\geqslant0}(-)^k\binom{n}{k}w^k z^{n-k}\,, \ \
(z-w)_{z}^{n} \doteq \sum_{k\geqslant0}(-)^{n+k}\binom{n}{k}w^{n-k}z^k
\end{equation}
which indicates, in positive powers of what variable ($w$ or $z$ in this
case) the expansion is performed. So, e.g., $(z-w)_{w}^{n}$ is analogous to
$(z-w)_{|z|>|w|}^{n}$ in terms of complex plane.

Of crucial importance in the present formalism is the $\delta$-function:
\begin{equation}
\label{delta}
\delta(z-w) = \delta(w-z) = \sum_{n}w^n z^{-n-1}
 = \ldots +\frac{z}{w^2}+\frac{1}{w}+\frac{1}{z}+\frac{w}{z^2}+\ldots\,.
\end{equation}
It exhibits standard-looking properties
\begin{equation}
\label{}
f(z)\,\delta(z-w) = f(w)\,\delta(z-w)\,, \ \
(z-w)\,\delta(z-w) = 0\,, \ \
\text{Res}_z f(z)\,\delta(z-w) = f(w)
\end{equation}
where $\text{Res}_z$ stands for a coefficient of $z^{-1}$\,,
being an analog of $(2\pi i)^{-1}\!\oint\!dz$ on a complex plane.
Note $\text{Res}_z\!\circ\partial_z = 0$\,.
Evidently, the definition (\ref{delta}) mimics the well-known
formulas like
$\delta(x-y) = \sum_n e^{in(x-y)} = \sum_n (e^{ix})^n (e^{iy})^{-n}$\,.

Useful relations with the derivatives of the $\delta$-function
are listed below:
\begin{gather}
\label{deriv}
\frac{1}{n!}\,\partial_{w}^{n}\,\delta(z-w)
 = \frac{(-)^n}{n!}\,\partial_{z}^{n}\,\delta(z-w)
 = \sum_{m}\binom{m}{n}w^{m-n}z^{-m-1} \ \ \ \ \ \ \
(n\geqslant0) \\
(z-w)^m\,\partial_{w}^{n}\,\delta(z-w) = 0 \ \ \ \ \ \ (m>n\geqslant0) \\
(z-w)^m\frac{1}{n!}\,\partial_{w}^{n}\,\delta(z-w)
 = \frac{1}{(n\!-\!m)!}\,\partial_{w}^{n-m}\delta(z-w) \ \ \ \ \ \
(0\leqslant m\leqslant n)  \label{dd} \\
\text{Res}_z\,f(z)\,\partial_{w}^{n}\,\delta(z-w)
 = \partial_{w}^{n}\,f(w)\,, \ \ \
\text{Res}_z\,\frac{(z-w)^m}{n!}\,\partial_{w}^{n}\,\delta(z-w)
 = \delta_{m,n} \ \ \  (m\,,n\geqslant0) \label{rd}
\end{gather}
The following formula may also be of some use:
\begin{equation}
\label{grenze}
(z-w)_w^{-n}-(z-w)_z^{-n}
 = \frac{1}{(n-1)!}\,\partial_{w}^{n-1}\delta(z-w)
\ \ \ \ \ \ (n>0)\,.
\end{equation}

Another principal notion here is locality. For the (most
important) case of two variables the locality condition is
\begin{equation}
\label{local}
(z-w)^n a(z,w)=0 \ \ \ \ \ \ \ (n\geqslant N\geqslant0)\,.
\end{equation}
It imposes severe restrictions on $a(z,w)$\,. Consider first the
simplest case
\begin{gather}
(z-w)\,b(z,w)=0 \ \ \ \ \ \ \ \Rightarrow \ \ \ \ \ \ \
b_{m+1,n}=b_{m,n+1} \notag\\
\ \ \ \ \ \ \ \Rightarrow \ \ \ \ \
b(z,w) = c(w)\,\delta(z-w)\,, \ \
c(w) = \text{Res}_z b(z,w)\,.
\label{loc1}
\end{gather}
If now only $(z-w)^2 a(z,w)=0$\,, we consecutively find that
\begin{gather*}
(z-w)\,a(z,w)=c_1(w)\,\delta(z-w)
\equiv c_1(w)\,(z-w)\,\partial_w\delta(z-w)\,, \\
c_1(w) = \text{Res}_z (z-w)\,a(z,w)\,, \\
a(z,w) - c_1(w)\,\partial_w\delta(z-w) = c_0(w)\,\delta(z-w)\,, \\
c_0(w) = \text{Res}_z [a(z,w)-c_1(w)\,\partial_w\delta(z-w)]
 = \text{Res}_z a(z,w)
\end{gather*}
and, therefore,
\begin{equation}
\label{loc2}
a(z,w)=c_0(w)\,\delta(z-w)+c_1(w)\,\partial_w\delta(z-w)\,.
\end{equation}
By induction, using (\ref{dd}) and (\ref{rd})\,, we show that
$(z-w)^n a(z,w)=0$ entails
\begin{equation}
\label{locn}
a(z,w)=\sum_{m=0}^{n-1}c_m(w)\frac{1}{m!}\partial_{w}^{m}\delta(z-w)\,,
\ \ \ \ c_m(w)=\text{Res}_z (z-w)^m a(z,w)\,.
\end{equation}
It is easily seen from (\ref{loc1}) or (\ref{locn}) that
a local formal distribution should contain infinitely many nonzero
terms of positive and negative degree in both variables, otherwise
it is identically zero. It is also clear that $a(w,z),\,
\partial_z a(z,w), \,\partial_w a(z,w),\,
f(z)\,a(z,w)$ and $f(w)\,a(z,w)$ are local if $a(z,w)$ is.

One interesting example of a local distribution is provided
by the product of two series with all unity coefficients:
\begin{equation}
\label{delta1}
(\sum_n z^{-n-1})(\sum_m w^{-m-1}) = \delta(z-1)\,\delta(w-1)
= \delta(z-w)\,\delta(w-1)
\end{equation}
(setting one argument of $\delta$-function to a nonzero number
makes sense)\,.

One can verify (directly, or by induction) the following analog of
the Taylor formula:
\begin{equation}
\label{Taylor}
[f(z) - \sum_{n=0}^{N}\,\frac{(z-w)^n}{n!}\,\partial_{w}^{n}\,f(w)\,]
\,\partial_{w}^{N}\,\delta(z-w) = 0\,.
\end{equation}
It is precisely in this way we may consider an expression inside square
brackets in (\ref{Taylor}) as something of the order $(z-w)^{N+1}$\,.

The following formulas with binomial coefficients are widely used
in the calculations performed (or implicitly meant) in the present text:
\begin{gather}
\binom{-n-1}{m} = (-)^m \binom{n+m}{m} \ \ \ \ \ (m\geqslant0)\,,
\ \ \ \ \ \ \ \ \binom{n}{m}=0 \ \ \ \ \ (m>n\geqslant0)\,, \\
\sum_{k=m}^{n}(-)^k \binom{n}{k}\binom{k}{m}=(-)^n\delta_{m,n}\,,
\ \ \ \  \sum_{k=0}^{n}(-)^k k^m\binom{n}{k}=(-)^n\,n!\,\delta_{m,n}\,,
\ \ \ \ (n\geqslant m\geqslant0) \\
\sum_{k=0}^{n}(-)^k \binom{n}{k}\binom{p-k}{m}=\binom{p-n}{m-n}\,,
\ \ \ \ \ \ \
\sum_{k=0}^{n}(-)^k \binom{n}{k}\binom{p+k}{m}=(-)^n\binom{p}{m-n}
\end{gather}

\vspace{.3cm}

\begin{center}
\large\textbf{Operator product expansion}
\end{center}

\vspace{.1cm}

A single-variable formal distribution $a(z)$ is called a field
on a vector space $V$ if $a_n$ are operators on $V$\,, and for
any $v\in V$ a series
$a(z)\,v$ contains only finitely many negative powers of $z$\,.
In other words, $a_n v = 0$ for $n\geqslant N(v)$\,.
Fields $a$ and $b$ are called (mutually) local if for some
$N(a,b)\geqslant0$
\begin{equation}
\label{locf}
(z-w)^N [a(z),b(w)]=0 \ .
\end{equation}
In what follows, we assume all fields to be pairwise local.

For each $n\in\mathbb{Z}$\,, an important notion of $n$-product of
two fields $a$ and $b$ is introduced as follows:
\begin{equation}
\label{nprod}
(a_n b)(w) = \sum_{m}\frac{(a_n b)_m}{w^{m+1}}\,, \ \ \ \
(a_n b)_m \doteq \sum_{k\geqslant0}(-)^k\binom{n}{k}
(a_{n-k}b_{m+k} -(-)^n b_{m+n-k}a_k)\,.
\end{equation}
It is seen from (\ref{nprod}) that $(a_n b)(w)$ is also a local field,
because for given $n$ and $v$ (or $c(z)$) we are able to find such $N$
that $(a_n b)_m v = 0$ (resp. $(a_n b)_m c = 0$) for $m\geqslant N$\,.

Let us discuss the definition of the $n$-product in more details.
For $n\geqslant0$ it is equal to
\begin{equation}
\label{n>0}
(a_n b)(w)=\text{Res}_z (z-w)^n [a(z),b(w)]\,.
\end{equation}
Locality condition (\ref{locf}) implies $(a_n b)(w)=0 \ (n\geqslant N)$
and, due to (\ref{locn}),
\begin{equation}
\label{commf}
[a(z),b(w)]=\sum_{n=0}^{N-1}(a_n b)(w)\frac{1}{n!}
\partial_{w}^{n}\delta(z-w)\,.
\end{equation}
Therefore, a commutator of local fields and (a finite number of)
their $n$-products with non-negative $n$ are closely related.
This is also seen at the component level ($n\geqslant0$)\,:
\begin{equation}
\label{commm}
(a_n b)_m = \sum_{k=0}^{n}(-)^k\binom{n}{k}[a_{n-k},b_{m+k}]\,, \ \ \ \ \
[a_p,\,b_q] = \sum_{k\geqslant0}\binom{p}{k}(a_k b)_{p+q-k}
\end{equation}
(the second sum is also finite due to locality)\,.

A special case $n=-1$ corresponds to the normal (or normally ordered)
product,
\begin{equation}
\label{normw}
(a_{-1}b)_m = \sum_{k<0}a_k b_{m-k-1} + \sum_{k\geqslant0}b_{m-k-1}a_k
\ \ \ \Rightarrow \ \ \
(a_{-1}b)(w) = \ :\!a(w)b(w)\!:\, \ \equiv \,\ :\!ab\!:\!(w)\,,
\end{equation}
where
\begin{equation}
\label{normzw}
:\!a(z)b(w)\!:\, \doteq a_+(z)b(w)+b(w)a_-(z)\,, \ \ \
a_+(z) = \sum_{n<0}\frac{a_n}{z^{n+1}}\,, \ \ \
a_-(z) = \sum_{n\geqslant0}\frac{a_n}{z^{n+1}}\ .
\end{equation}
Usually $a_+(z)$ is said to contain creation and $a_-(z)$
annihilation operators. We can now verify that the following general
formula for negative $n$ agrees with the definition (\ref{nprod})\,:
\begin{equation}
\label{n<0}
(a_n b)(w) = \ \frac{1}{(-n-1)!}:\!\partial^{-n-1}a(w)\cdot b(w)\!:
\ \ \ \ \ \ \ \ (n<0)
\end{equation}
There also exists a universal form of the $n$-product definition:
\begin{equation}
\label{resprod}
(a_n b)(w)=\text{Res}_z\bigl(a(z)b(w)(z-w)^n_w-b(w)a(z)(z-w)^n_z\bigr)\,.
\end{equation}

Remarkably, $n$-products play the role of coefficient functions of the
operator product expansion (OPE). In CFT${}_2$\,, the latter is formulated for
the radially ordered products
\begin{equation}
\label{rad}
\mathcal{R}\,a(z)\,b(w) \doteq
\begin{cases}
a(z)\,b(w) & \qquad |z|>|w| \\ b(w)\,a(z) & \qquad |w|>|z|
\end{cases}
\end{equation}
Within the present formalism, OPE relations are expressed
in terms of the definitions (\ref{expanzw})\,:
\begin{equation}
\label{ope}
a(z)\,b(w) = \sum_n \frac{(a_n b)(w)}{(z-w)_{w}^{n+1}}\ , \ \ \ \
b(w)\,a(z) = \sum_n \frac{(a_n b)(w)}{(z-w)_{z}^{n+1}}
\end{equation}
or, equivalently,
\begin{gather}
\label{opew}
a(z)\,b(w) = \sum_{n=0}^{N-1} \frac{(a_n b)(w)}{(z-w)_{w}^{n+1}}
\ + :\!a(z)\,b(w)\!:\, \\
\label{opez}
b(w)\,a(z) = \sum_{n=0}^{N-1} \frac{(a_n b)(w)}{(z-w)_{z}^{n+1}}
\ + :\!a(z)\,b(w)\!:\, \\
\label{ope<0}
:\!a(z)\,b(w)\!:\, = \sum_{n\geqslant0}(a_{-n-1}b)(w)\ (z-w)^n
\end{gather}
Eqs. (\ref{ope<0}) and (\ref{ope}) are to be read not literally,
but in the sense of Taylor's expansion (\ref{Taylor}),
its coefficients given by (\ref{n<0})\,.
At the same time, (\ref{opew}) and (\ref{opez}) are derived from
(\ref{grenze}) and
\begin{gather}
a(z)\,b(w) \ - :\!a(z)\,b(w)\!:\ =\, [a_-(z),b(w)]\,, \\
b(w)\,a(z) \ - :\!a(z)\,b(w)\!:\ =\, [b(w),a_+(z)]
\end{gather}
by extracting negative and non-negative
powers, respectively, from the $z$-expansion of (\ref{commf})\,,
and are identically true. We see that a singular part of OPE is given
by $n$-products with non-negative $n$\,, or by the commutator of fields.

It seems to be instructive to rewrite this derivation of OPE using the
CFT${}_2$ language. Let
$\mathcal{R}\,a(z)\,b(w) = \sum_n \varphi_n(w)(z-w)^{-n-1}$\,.
Then
\begin{multline}
\label{derivope}
\varphi_n(w) = \text{Res}_{z-w}\bigl((z-w)^n\,\mathcal{R}\,a(z)\,b(w)\bigr)\\
= \frac{1}{2\pi i}\oint_w dz (z-w)^n\,\mathcal{R}\,a(z)\,b(w)
= \frac{1}{2\pi i}\left(\oint_{|z|>|w|}dz - \oint_{|z|<|w|}dz\right)
 (z-w)^n\,\mathcal{R}\,a(z)\,b(w) \\
= \frac{1}{2\pi i}\oint dz
 \left(a(z)\,b(w)\,(z-w)^n_{|z|>|w|}-b(w)\,a(z)\,(z-w)^n_{|z|<|w|}\right) \\
= \text{Res}_z
 \left(a(z)\,b(w)\,(z-w)^n_{|z|>|w|}-b(w)\,a(z)\,(z-w)^n_{|z|<|w|}\right)
= (a_n b)(w)\,.
\end{multline}

It should be noted that from (\ref{nprod}) and (\ref{commm})
(i.e., from the definition of the $n$-product plus the locality condition),
the following useful formula (Borcherds identity) can be deduced. Namely,
for $l,m,n\in\mathbb{Z}$\,, and $a,b,c$ being fields, the following relations
hold:
\begin{equation}
\label{Borch}
\sum_{k\geqslant0}\binom{m}{k}(a_{l+k}b)_{m+n-k}c
= \sum_{k\geqslant0}(-)^k\binom{l}{k}
\bigl(a_{m+l-k}(b_{n+k}c) -(-)^l b_{n+l-k}(a_{m+k}c)\bigr)\,.
\end{equation}

In CFT${}_2$\,, it's common practice to construct composite operators
as normal products of two (or more) fields. To find then the OPE relations
with these composite fields, the so-called Wick formulas are used. In the
present formalism the latter are obtained from (\ref{Borch})\,,
or (\ref{commm})\,, and look like
\begin{equation}
\label{Wick}
a_n(b_{-1}c) = b_{-1}(a_n c)
 + \sum_{k\geqslant0}\binom{n}{k}(a_k b)_{n-k-1}c\ .
\end{equation}

\vspace{.3cm}

\begin{center}
\large\textbf{Vertex algebras}
\end{center}

\vspace{.1cm}

A vertex algebra consists of a vector space $V$ (whose elements
are called `states'), a vacuum state $\Omega\in V$\,, and a
`state-field correspondence' linear map $Y$ which
associates a field $a(z)\equiv Y(a,z)=\sum_n a_n z^{-n-1}$
(vertex operator) to each
$a\in V$ in such a way that $Y(a_n b,z)=(a_nb)(z)$
are given by (\ref{nprod}), all fields are mutually local
($a_nb$\,=\,0 for $n\geqslant N(a,b)$)\,, and $Y(\Omega,z)$ is equal
to unity ($\Omega_n=\delta_{n,-1}\,\mathbf{1}$)\,. In addition,
a shift operator $D$ is defined on $V$ by
$(Da)(z) \doteq \partial_z a(z)$\,.
Note that an expression $a_nb$ is used here in two meanings:
it denotes a state (an element in $V$ obtained by the action of
the operator $a_n$ upon $b\in V$), and serves simultaneously as a name of
the field $(a_nb)(z)$ (associated to this state via the map $Y$)\,,
which is exactly the $n$-product of two fields $a(z)$ and $b(z)$\,.

Now let us study the consequences of the above definitions.
Due to locality of vertex operators, we may use not only
(\ref{nprod})\,, but also Borcherds' relation (\ref{Borch})\,. Evidently,
\begin{equation}
\label{vac}
a_{-1}\,\Omega=a\,, \ \ \ \ \ \
a_n\,\Omega=0 \ \ \ (n\geqslant0)\,, \ \ \ \ \ \ \
D\,\Omega=0\,, \ \ \ \ \ \
(Da)_n=-na_{n-1}\,.
\end{equation}
Further, for $k\geqslant0$\,,
\begin{equation}
\label{exp}
(D^k a)_n = (-)^k\,k!\,\binom{n}{k}\,a_{n-k}\,, \ \ \ \
a_{-k-1} = \frac{1}{k!}\,(D^k a)_{-1} \ \ \ \ \ \
\Rightarrow \ \ \ \ Y(a,z)\,\Omega = e^{zD}a\,.
\end{equation}
Using (\ref{Borch}) with $m=0, \,n=-2, \, c=\Omega$\,, we obtain
$D(a_lb)=(Da)_lb+a_l(Db)\,,$ whence
\begin{equation}
\label{Leib}
[D,\,a_n]=(Da)_n=-na_{n-1} \ \ \ \ \ \ \Rightarrow \ \ \ \
[D,\,Y(a,z)]=Y(Da,z)=\partial_z Y(a,z)\,.
\end{equation}
As an immediate generalization we find
\begin{equation}
\label{transl}
e^{zD}Y(a,w)e^{-zD} = Y(e^{zD}\!a,w) = Y(a,w+z)_z\ .
\end{equation}
For $m=-1, \,n=0, \, c=\Omega$ \ formula (\ref{Borch}) yields
a so-called skew symmetry property
\begin{equation}
\label{skew}
b_la = \sum_{k\geqslant0}\,\frac{(-)^{l+k+1}}{k!}\,D^k(a_{l+k}b)
\ \ \ \ \ \ \Rightarrow \ \ \ \
Y(b,z)\,a = e^{zD}\,Y(a,-z)\,b\,.
\end{equation}
The OPE relation assumes here the following form:
\begin{equation}
\label{opey}
Y(a,z)Y(b,w) = Y(Y(a,z-w)_w\,b,w)\,, \ \ \ \
Y(b,w)Y(a,z) = Y(Y(a,z-w)_z b,w)\,.
\end{equation}

If $V={\scriptstyle\bigoplus_{k\geqslant0}}V_k$ is a graded space
equipped with a vertex algebra structure (so that $\Omega\in V_0\,,
\ D:V_k\rightarrow V_{k+1}$\,, \,all $a_n$ homogeneous),
one usually renumbers the components of the field $a(z)$ with
$a\in V_\Delta$ in such a way (shifting index $n\rightarrow n-\Delta+1$)
that
\begin{equation}
\label{grad}
a(z)=\sum_n\,a_n\,z^{-n-\Delta}, \ \ \
a_{-n}:V_k\rightarrow V_{k+n}\,, \ \ \
a_{-\Delta}\Omega= a\,, \ \ \
a_n\Omega=0 \ \ (n>-\Delta)\,.
\end{equation}
As the famous example, the Virasoro algebra with central charge $c$
can be reconstructed in this way from a single `conformal' vector
$\omega\in\!V_2$\ :
\begin{gather}
\label{Vir}
\omega(z)\,\equiv\, Y(\omega,z)=\sum_n\,\omega_n\,z^{-n-1} \
\equiv \,T(z) =\sum_n\,L_n\,z^{-n-2} \ \ \ \ \ (\omega_n=L_{n-1})\ , \\
\label{Vir1}
L_{-2}\,\Omega=\omega\,, \ \ \
L_{-1}=D\,, \ \ \
L_0\,a = \Delta\,a \ \ \,(a\in V_{\Delta})\,, \ \ \
L_1\,\omega=0\,, \ \ \
L_2\,\omega=\frac{c}{2}\ \Omega\,.
\end{gather}
These definitions prove to be consistent, agree with (\ref{Borch})\,,
and ultimately result in
\begin{gather}
[L_m,L_n] =
 (m-n)L_{m+n}+\frac{c}{12}\,m(m^2-1)\,\delta_{n,-m} \ , \\
\mathcal{R}\,T(z)\,T(w) = \frac{c}{2}\,(z-w)^{-4}+2\,(z-w)^{-2}\,T(w)
 +(z-w)^{-1}\,\partial\, T(w)+\,\ldots\ .
\end{gather}

\vspace{.3cm}

\begin{center}
\large\textbf{Free fermions}
\end{center}

\vspace{.1cm}

Below we describe the (Sugawara-like) construction of a conformal vector
as a bilinear combination of `more elementary' objects: in this (simplest)
case, neutral fermions. The corresponding vertex algebra is generated by the
vacuum state $\Omega$ and one more vector $\phi$ subject to
\begin{equation}
\label{neutral}
\phi_0\phi = \Omega\,, \ \ \ \ \ \
\phi_n\phi = 0 \ \ \ \ \ (n>0)\ .
\end{equation}
This exactly corresponds to
\begin{equation}
\label{phiz}
\phi(z) = \sum_n\,\phi_n\,z^{-n-1}\,, \ \ \ \ \
[\phi_m,\phi_n]_+ = \delta_{m+n,-1}\ , \ \ \ \ \
\phi(z)\,\phi(w) \sim \frac{1}{(z-w)_w}\ .
\end{equation}
Since $\phi(z)$ is a fermionic field, we deal with anticommutators
instead of commutators, and all formulas like (\ref{nprod}), (\ref{commm}),
and so on, should be modified accordingly.

Let us now introduce a vector
\begin{equation}
\label{omff}
\omega \doteq \frac{1}{2}\,\phi_{-2}\phi
 \equiv \frac{1}{2}\,(D\phi)_{-1}\phi
 \equiv \frac{1}{2}\,\phi_{-2}\phi_{-1}\Omega
\end{equation}
which is to be recognized as a conformal one. Eq.\;(\ref{Wick}) yields
\begin{equation}
\label{}
\phi_0\omega = -\frac{1}{2}\,D\phi\,, \ \ \ \ \
\phi_1\omega = \frac{1}{2}\,\phi\,, \ \ \ \ \
\phi_n\omega = 0 \ \ \ \ \ (n>1)
\end{equation}
and (\ref{skew}) then gives
\begin{equation}
\label{omphi}
\omega_0\phi = D\phi\,, \ \ \ \ \
\omega_1\phi = \frac{1}{2}\,\phi\,, \ \ \ \ \
\omega_n\phi = 0 \ \ \ \ \ (n>1)
\end{equation}
Analogously, we come to
\begin{equation}
\label{}
\omega_0 D\phi = D^2\phi\,, \ \ \ \ \
\omega_1 D\phi = \frac{3}{2}\,D\phi\,, \ \ \ \ \
\omega_2 D\phi = \phi\,,
\end{equation}
and, finally, to
\begin{equation}
\label{omom}
\omega_0\omega = D\omega\,, \ \ \ \ \
\omega_1\omega = 2\omega\,, \ \ \ \ \
\omega_3\omega = \frac{1}{4}\,\Omega
\end{equation}
(only nonzero terms are displayed). So $\omega$ (\ref{omff})
is a genuine conformal vector, with central charge $c=1/2$\,,
and $\phi$ a \textit{primary} element of conformal weight $\Delta=1/2$
(that means literally (\ref{omphi}))\,,
with the OPE as follows:
\begin{gather}
\label{OPEomf}
\omega(z)\,\phi(w) = \frac{\phi(w)}{2(z-w)^2_w}
 + \frac{\partial\phi(w)}{(z-w)_w} + \ldots\ , \\
\label{OPEomom}
\omega(z)\,\omega(w) = \frac{1}{4(z-w)^4_w}
 + \frac{2\omega(w)}{(z-w)^2_w}
 + \frac{\partial\omega(w)}{(z-w)_w}\,+\,\ldots\ .
\end{gather}
Note that in this case, unlike (\ref{grad}) and (\ref{Vir}),
we prefer not to shift indices of the field components in (\ref{phiz}).

As for the vertex algebra as a whole, it is spanned, according to
the Pauli principle, by the (finite) monomials of the form
\begin{equation}
\label{span}
\ldots \phi_{-m}^{i_m} \ldots \phi_{-1}^{i_1}\Omega\ \ \ \ \ \
(m>0, \ \ i_m = 0,1)\ .
\end{equation}
The second equality in (\ref{commm}) gives
\begin{equation}
\label{}
[\omega_0,\phi_n] = (D\phi)_n = -n\phi_{n-1}\,, \ \ \ \ \
[\omega_1,\phi_n] = (-\frac{1}{2}-n)\phi_n\,,
\end{equation}
and one readily verifies that
\begin{equation}
\label{omgen}
\omega_0 a = Da\,, \ \ \ \ \ \ \ \omega_1 a = \Delta a
\end{equation}
for any $a$ of the type (\ref{span})\,, in accordance with the definition
(\ref{Vir1}) of conformal vector.

\vspace{.3cm}

Let us now proceed with a slightly more involved example related to charged
fermions. Here the vertex algebra is generated by the vacuum state $\Omega$
and two fermionic (odd) states $\psi^+\ (\equiv \psi)$
and $\psi^-\ (\equiv \bar\psi)$ subject to
\begin{equation}
\label{charged}
\psi^+_0\psi^- = \psi^-_0\psi^+ = \Omega\,, \ \ \ \ \ \ \
\psi^+_0\psi^+ = \psi^-_0\psi^- =
\psi^{\pm}_n\psi^{\pm} = \psi^{\pm}_n\psi^{\mp} = 0 \ \ \ \ (n>0)
\end{equation}
or, in other words (note renumbering $\psi_n^+$ w.r.t.\,(\ref{psi0})\,)\,,
\begin{gather}
\label{psiz}
\psi^{\pm}(z) = \sum_n\,\psi^{\pm}_n\,z^{-n-1}\,, \ \ \ \ \
[\psi^{\pm}_m,\psi^{\pm}_n]_+ = 0\,, \ \ \ \ \
[\psi^{\pm}_m,\psi^{\mp}_n]_+ = \delta_{m+n,-1}\ , \\
\label{psiOPE}
\psi^{\pm}(z)\,\psi^{\pm}(w) \sim 0\,, \ \ \ \ \ \
\psi^+(z)\,\psi^-(w) \sim \psi^-(z)\,\psi^+(w) \sim \frac{1}{(z-w)_w}\ .
\end{gather}
Now we can construct a bilinear (bosonic) current (\ref{current})
\begin{equation}
\label{J}
J = \psi^+_{-1}\psi^- = \psi^+_{-1}\psi^-_{-1}\Omega
\end{equation}
and, for arbitrary complex number $\lambda$\,,
a conformal vector
\begin{equation}
\label{omlam}
\omega_\lambda = \frac{1}{2}J_{-1}J + (\frac{1}{2}-\lambda)DJ
 = (1-\lambda)\,\psi^+_{-2}\psi^- + \lambda\,\psi^-_{-2}\psi^+
 = [(1-\lambda)\,\psi^+_{-2}\psi^-_{-1}
 + \lambda\,\psi^-_{-2}\psi^+_{-1}]\Omega\,.
\end{equation}
We will see that $\psi^+$ and $\psi^-$ are primary states of conformal weights
$\lambda$ and $1-\lambda$\,, respectively, whereas $J$ has $\Delta=1$ and
becomes primary only for $\lambda=1/2$ (when $c=1$).

First, we check that two definitions of $\omega_\lambda$ (in terms of $J$ and
$\psi^{\pm}$) do agree. Using (\ref{nprod}), (\ref{psiz}), and properties of
$D$\,, we find
$$
J_{-1}J = (\psi^+_{-1}\psi^-)_{-1}J
 = \sum_{k\geqslant 0}(\psi^+_{-1-k}\psi^-_{-1+k} - \psi^-_{-2-k}\psi^+_k)
  \,\psi^+_{-1}\psi^-_{-1}\Omega
 = (\psi^+_{-2}\psi^-_{-1} + \psi^-_{-2}\psi^+_{-1})\,\Omega\,,
$$
$$
DJ = (D\psi^+)_{-1}\psi^- + \psi^+_{-1}D\psi^-
 = (\psi^+_{-2}\psi^-_{-1} + \psi^+_{-1}\psi^-_{-2})\,\Omega
 = (\psi^+_{-2}\psi^-_{-1} - \psi^-_{-2}\psi^+_{-1})\,\Omega\,,
$$
which verifies (\ref{omlam})\,.

Now we are ready to obtain, quite straightforwardly, the $n$-products
(with $n\geqslant0$), (anti)commutators, and singular parts of the OPE
for $\psi^{\pm}, J$ and $\omega_\lambda$ in all the relevant combinations
(as usual, only nonzero contributions are listed)\,:
\begin{gather}
\label{Jnpsi}
\psi_0^{\pm}J = \mp \psi^{\pm}\,, \ \ \ \ \
J_0\psi^{\pm} = \pm \psi^{\pm}\,, \ \ \ \ \
[J_m,\psi_n^{\pm}] = \pm\psi^{\pm}_{m+n}\,, \\
\label{Jzpsiw}
\psi^{\pm}(z)J(w) \sim \mp\frac{\psi^{\pm}(w)}{(z-w)_w}\,, \ \ \ \ \
J(z)\,\psi^{\pm}(w) \sim \pm\frac{\psi^{\pm}(w)}{(z-w)_w}\,, \\
\label{JJ}
J_1 J = \Omega\,, \ \ \ \ [J_m,J_n] = m\delta_{m,-n}\,, \ \ \ \
J(z)J(w) \sim \frac{1}{(z-w)^2_w}\,, \\
\label{ompsi}
\omega_\lambda{}_0 \psi^{\pm} = D\psi^{\pm}\,, \ \ \ \ \
\omega_\lambda{}_1 \psi^+ = \lambda\psi^+\,, \ \ \ \ \
\omega_\lambda{}_1 \psi^- = (1-\lambda)\,\psi^-\,, \\
\label{ompsiOPE}
\omega_\lambda(z)\,\psi^{\pm}(w) \sim
\frac{[1\pm(2\lambda-1)]\psi^{\pm}(w)}{2(z-w)^2_w}
 + \frac{\partial\psi^{\pm}(w)}{(z-w)_w}\,, \\
\label{Jom}
J_1\omega_\lambda = J\,, \ \ \ \
J_2\omega_\lambda = (1-2\lambda)\,\Omega\,, \ \ \ \
J(z)\,\omega_\lambda(w) \sim \frac{1-2\lambda}{(z-w)^3_w}
 + \frac{J(w)}{(z-w)^2_w}\,, \\
\label{omJ}
\omega_\lambda{}_0 J = DJ\,, \ \ \ \ \
\omega_\lambda{}_1 J = J\,, \ \ \ \ \
\omega_\lambda{}_2 J = (2\lambda-1)\,\Omega\,, \\
\label{omJOPE}
\omega_\lambda(z)J(w) \sim \frac{2\lambda-1}{(z-w)^3_w}
 + \frac{J(w)}{(z-w)^2_w} + \frac{\partial J(w)}{(z-w)_w}\,, \\
\label{omlaomla}
\omega_\lambda{}_0\omega_\lambda = D\omega_\lambda\,, \ \ \ \ \
\omega_\lambda{}_1\omega_\lambda = 2\omega_\lambda\,, \ \ \ \ \
\omega_\lambda{}_3\omega_\lambda = (-6\lambda^2+6\lambda-1)\,\Omega\,, \\
\label{omomOPE}
\omega_\lambda(z)\,\omega_\lambda(w) \sim
 \frac{-6\lambda^2+6\lambda-1}{(z-w)^4_w}
 + \frac{2\omega_\lambda(w)}{(z-w)^2_w}
 + \frac{\partial\omega_\lambda(w)}{(z-w)_w}\,,
\end{gather}
the central charge here being \,$c=-12\lambda^2+12\lambda-2
=1-3(2\lambda-1)^2$\,.

Let us now consider an important set of primary states $V^m$
(the same as $|\,m\rangle$ in (\ref{mvac0})\,)\,:
\begin{equation}
\label{mvac}
V^m = \psi_{-m}^{+}\psi_{-m+1}^{+}\ldots\psi_{-1}^{+}\Omega\,, \ \ \ \ \
V^{-m} = \psi_{-m}^{-}\psi_{-m+1}^{-}\ldots\psi_{-1}^{-}\Omega \ \ \ \ \ \
(m\geqslant 0)\,.
\end{equation}
Of course, \,$V^0 = \Omega\,, \ V^{\pm 1}=\psi^{\pm}$\,.
Using (\ref{psiz}) and (\ref{Jnpsi}), one easily checks that
\begin{equation}
\label{Jmvac}
J_{n>0}V^m = 0\,, \ \ \ \ \ J_0 V^m = m V^m\,, \ \ \ \ \
m J_{-1}\!V^m = D V^m\,.
\end{equation}
The last equality is verified by induction (say, for $m>0$)\,:
\begin{gather}
\notag
J_{-1}V^1 = J_{-1}\psi_{-1}^+\Omega = \psi_{-2}^+\Omega + \psi_{-1}^+ J
 = (D\psi^+)_{-1}\Omega + \psi_{-1}^+\psi_{-1}^+\psi^- = DV^1, \\
\notag
\psi_{-m-1}^+ J_{-1}V^m = 0\,, \ \ \ \
J_{-1}V^{m+1} = \psi_{-m-2}^+ V^m = \frac{1}{m+1}[D,\psi_{-m-1}^+]V^m
 = \frac{D V^{m+1}}{m+1}\,.
\end{gather}
Now, from
\begin{equation}
\label{}
\omega_{\lambda n} = \frac{1}{2}\,\left(\,\sum_{k<0} J_k J_{n-k-1} +
\sum_{k\geqslant 0} J_{n-k-1}J_k\right) + (\lambda
 - \frac{1}{2})\,n J_{n-1}
\end{equation}
and, in particular,
\begin{equation}
\label{}
\omega_{\lambda 0} = \sum_{k\geqslant 0} J_{-k-1}J_k\,, \ \ \ \ \
\omega_{\lambda 1} = \frac{1}{2}\,J_0 J_0 + (\lambda - \frac{1}{2})J_0
 + \sum_{k>0} J_{-k}J_k\,,
\end{equation}
we see that
\begin{equation}
\label{primV}
\omega_{\lambda 0}V^m = J_{-1}J_0 V^m = m J_{-1}V^m = DV^m\,, \ \ \ \ \
\omega_{\lambda 1}V^m = \left(\frac{m^2}{2} + m\lambda
 - \frac{m}{2}\right)V^m\,.
\end{equation}
These relations, together with $\omega_{\lambda n}V^m = 0$ for $n\!>\!1$,
characterize $V^m$ as a primary state of conformal weight
$(\frac{m^2}{2}+m\lambda-\frac{m}{2})$\,.

There exists a remarkable representation for the field $V^m(z)$ in terms of
$J_n$ and an additional invertible operator $Y$ (previously introduced
in (\ref{K}); not to be confused with the
state-field map itself!) defined by
\begin{equation}
\label{Y}
Y\psi_n^+ = \psi_{n-1}^+Y\,, \ \ \ Y\psi_n^- = \psi_{n+1}^-Y\,,
\ \ \ Y\Omega = V^1 \ \ \Longrightarrow \ \ \ YV^m = V^{m+1},
\ \ \ V^m = Y^m\Omega\,.
\end{equation}
From
\begin{equation}
\label{}
J_{n\neq0} = \sum_k \psi_k^+ \psi_{n-k-1}^-\,, \ \ \ \ \ \
J_0 = \sum_{k\geqslant0} (\psi_{-k-1}^+\psi_k^- - \psi_{-k-1}^-\psi_k^+)
\end{equation}
one immediately concludes that
\begin{equation}
\label{YJ}
J_n Y = Y J_n \ \ (n\neq0)\,, \ \ \ \ \ \ [J_0,Y] = Y \ \ \ \ \
\Longrightarrow \ \ \ \ \ [J_0,Y^m] = m Y^m\,.
\end{equation}
A lengthy argument based on the Schur lemma shows that
\begin{equation}
\label{Vz}
V^m(z) = \ :\!e^{m\varphi(z)}\!: \ = Y^m \exp(-m\sum_{n<0}\frac{J_n}{nz^n})
\ z^{mJ_0}\, \exp(-m\sum_{n>0}\frac{J_n}{nz^n})
\end{equation}
(cf. (\ref{psiF})\,,(\ref{vert})\,)\,, where
\begin{equation}
\label{phi}
\varphi(z) = \ln Y + J_0\ln z - \sum_{n\neq0}\frac{J_n}{nz^n}\,, \ \ \ \ \ \
\varphi\,'(z) = J(z)\,.
\end{equation}
We treat $\ln Y$ as a creation operator in accordance with
$[J_0,\ln Y]=1$\,. Differentiating $V^m(z)$ is in agreement with
(\ref{Jmvac})\,: \ $\partial_z V^m(z)= \ m:\!J(z)V^m(z)\!: \
\equiv m(J_{-1}V^m)(z)$\,. Another identity stemming from (\ref{Vz})\,,
$V^m(z)\,\Omega=e^{zD}V^m=\exp(-m\sum_{n<0}\frac{J_n}{nz^n})V^m$, is also
valid due to the properties of Schur polynomials.

\vspace{.3cm}

\begin{center}
\large\textbf{Virasoro algebra in KdV}
\end{center}

\vspace{.1cm}

The famous Korteweg\,-\,de\,Vries (KdV) equation
\begin{equation}
\label{KdV}
u_t = \frac{1}{4}\,(u_{xxx} + 6uu_x)
\end{equation}
is a (first non-trivial) member of an integrable hierarchy
\begin{equation}
\label{hierKdV}
L_{t_n} = [L_{+}^{(2n-1)/2},L] \ \ \ \ \ \ \ \ \ (t_1=x\,, \ t_2=t)
\end{equation}
generated by the Lax operator
\begin{equation}
\label{L}
L = \partial^2 + u\,, \ \ \ \ \ u = u(t,x)\,, \ \ \
\partial = \partial_x\,.
\end{equation}
The KdV equation corresponds to $n=2$\,, and so is governed by
the $M$-operator
\begin{equation}
\label{L3/2}
L_{+}^{3/2} = \partial^3 + \frac{3}{2}u\partial + \frac{3}{4}u_x
 = \partial^3 + \frac{3}{4}(u\partial + \partial\circ u)\,.
\end{equation}
The hierarchy (\ref{hierKdV}) may be also seen as a reduction of the
KP hierarchy which reduces its immense phase space to the (unique)
square root of $L$\,, with a single unknown function $u$\,:
\begin{equation}
\label{L1/2}
L^{1/2} = \partial + \frac{1}{2}u\partial^{-1} - \frac{1}{4}u_x\partial^{-2}
 + \frac{1}{8}(u_{xx}-u^2)\,\partial^{-3} + \mathcal{O}\,(\partial^{-4})\,.
\end{equation}

The following \textit{modified} KdV equation (mKdV) is closely related
to (\ref{KdV})\,:
\begin{equation}
\label{mKdV}
q_t = \frac{1}{4}\,(q_{xxx} - 6q^2 q_x)\,.
\end{equation}
Namely, it is obtained from (\ref{KdV}) by the Miura transformation
\begin{equation}
\label{Miura}
u = q_x - q^2 \ \ \ \ \ \Rightarrow \ \ \ \
L = \partial^2 + u = (\partial - q)\,(\partial + q)
\end{equation}
which sends (\ref{KdV}) to
$$
(\partial - 2q)\,(q_t - \frac{1}{4}q_{xxx} + \frac{3}{2}q^2 q_x) = 0\,.
$$

Let us now look at these integrable equations from the Hamiltonian standpoint.
Here
\begin{gather}
\label{brac}
f_t=\{H,f\}\,, \ \ \ \ \
\{F,G\}\doteq\int\!dx\,dy \frac{\delta F}{\delta u(x)}\{u(x),u(y)\}
 \frac{\delta G}{\delta u(y)}\,, \\
\label{var}
F(u+\varepsilon v)\doteq F(u)
 + \varepsilon\!\int\!dx\,v(x)\,\frac{\delta F}{\delta u(x)}
 + \mathcal{O}(\varepsilon^2)
\ \ \ \ \Rightarrow \ \ \ \
\frac{\delta F}{\delta u(x)}
 = \sum_{n=0}^{\infty}(-\partial)^n\frac{\partial F}{\partial u^{(n)}}\,,
\end{gather}
and $u^{(n)}\equiv \partial^n u$\,. Assuming locality of the Poisson bracket,
\begin{equation}
\label{u-brac}
\{u(x),u(y)\} = -\Lambda(x)\,\delta(x-y)\,,
\end{equation}
we come to
\begin{equation}
\label{brac2}
\{G,F\}=\int\!dx \frac{\delta F}{\delta u(x)}\Lambda(x)
 \frac{\delta G}{\delta u(x)}\,, \ \ \ \ \
u_t = \Lambda\frac{\delta H}{\delta u}\,.
\end{equation}

Hamiltonians $H_m$ of the KdV hierarchy are known to be of the form
$\sim\!\!\int dx\text{Res}L^{m+\frac{1}{2}}$ ( $\text{Res}=\text{Res}_\partial$
is a coefficient of $\partial^{-1}$)\,. For example,
\begin{equation}
\label{H012}
H_0 = \int\!dx\,u\,, \ \ \
H_1 = \frac{1}{4}\int\!dx\,u^2\,, \ \ \
H_2 = \frac{1}{16}\int\!dx\,(2u^3-u_x^2)\,.
\end{equation}
Moreover, KdV is a bi-Hamiltonian system, possessing (at least) two local
Poisson structures of the type (\ref{u-brac})\,:
\begin{equation}
\label{Lambda12}
\Lambda_1 = 2\partial\,, \ \ \ \ \ \
\Lambda_2 = \frac{1}{2}\partial^3 + 2u\partial + u_x
 = \frac{1}{2}\partial^3 + u\partial + \partial\circ u\,.
\end{equation}
Acting on adjacent Hamiltonians, these two $\Lambda$ operators produce
the same result:
\begin{gather}
u_x \equiv u_{t_1} = \Lambda_1\frac{\delta H_1}{\delta u}
 = \Lambda_2\frac{\delta H_0}{\delta u} = u_x \ \ \ \ \ \
(\text{identity}) \\
u_t \equiv u_{t_2} = \Lambda_1\frac{\delta H_2}{\delta u}
 = \Lambda_2\frac{\delta H_1}{\delta u} =  \frac{1}{4}(u_{xxx} + 6uu_x)
\ \ \ \ \ (\text{KdV}) \\
u_{t_m} = \Lambda_1\frac{\delta H_m}{\delta u}
 = \Lambda_2\frac{\delta H_{m-1}}{\delta u} = \ \
\text{$m$-th member of KdV hierarchy}
\end{gather}

It is the (symmetry) properties of the second Poisson bracket provided by
$\Lambda_2$ that appear to be the main subject of this section. To begin
with, we use the Miura transformation (\ref{Miura}) to discover that
\begin{equation}
\label{}
\Lambda_2 = \frac{1}{2}\partial^3 + 2u\partial + u_x
 = (\partial - 2q)\,(-\frac{\partial}{2})\,(-\partial - 2q)\,.
\end{equation}
Then, from (\ref{brac2}) and
$$
\frac{\delta u(x)}{\delta q(y)} = (\partial_x - 2q)\,\delta(x-y)\,,
\ \ \ \ \ \ \frac{\delta G}{\delta q(x)}
 = \int\!dy\,\frac{\delta u(y)}{\delta q(x)}\,\frac{\delta G}{\delta u(y)}
 = (-\partial_x - 2q)\,\frac{\delta G}{\delta u(x)}
$$
we see that $-\frac{1}{2}\partial$ is nothing but a Poisson structure of mKdV:
\begin{equation}
\label{q-brac}
\{q(x),q(y)\} = \frac{1}{2}\partial_x\delta(x-y)\,.
\end{equation}
Indeed, the second KdV bracket immediately follows from (\ref{q-brac})\,,
(\ref{Miura}) and (\ref{dd})\,. It is also readily verified that
\begin{equation}
\label{}
q_t = -\frac{1}{2}\partial\frac{\delta H_1(u(q))}{\delta q}
 = \frac{1}{4}(q_{xxx} - 6q^2q_x)\,.
\end{equation}

Our next observation is a striking similarity between the Poisson brackets
of (m)KdV and appropriate field commutators in CFT${}_2$\,. Let us rewrite
the second KdV bracket as
\begin{equation}
\label{u-brac2}
\{u(x),u(y)\} = -(\frac{c}{12}\partial_x^3
 +2u(x)\partial_x+u_x(x))\,\delta(x-y)\,, \ \ \ \ \ \ \ \ c=6\,.
\end{equation}
Compare (\ref{u-brac2}) and (\ref{q-brac}) with local commutators of the
energy-momentum tensor $T$ and free bosonic field $a$
(the same as $J$ (\ref{JJ})\,)\,, respectively:
\begin{align}
\label{TT}
 [T(z),\,T(w)] &= -(\frac{c}{12}\,\partial_z^3+2T(z)\,\partial_z
 +T_z(z))\,\delta(z-w)\,, \\
\label{aa}
[a(z),\,a(w)] &= -\partial_z\,\delta(z-w)\,.
\end{align}
Now it looks only natural to mimic the corresponding CFT${}_2$ definitions
as follows:
\begin{equation}
\label{defs}
u(x) = \sum_{n}x^{-n-2}L_n\,, \ \ \ \ \
q(x) = \sum_{n}x^{-n-1}a_n\,,
\end{equation}
to reproduce the Virasoro and Heisenberg algebras, up to inessential overall
factors:
\begin{align}
\label{-Vir}
\{L_m,L_n\} &= (m-n)L_{m+n} + \frac{c}{12}\,m(m^2-1)\,\delta_{n,-m}\,, \\
\label{aman}
\{a_m,a_n\} &= -\frac{m}{2}\,\delta_{n,-m}\,.
\end{align}
To recall the techniques, we show how to deduce (\ref{q-brac}) from
(\ref{aman}) and (\ref{deriv})\,:
\begin{multline}
\label{}
\{q(x),q(y)\} = \sum_{mn}x^{-m-1}y^{-n-1}\{a_m,a_n\}
 = -\frac{1}{2}\sum_m m x^{-m-1}y^{m-1} \\
= -\frac{1}{2}\partial_y\,\delta(y-x)
 = \frac{1}{2}\partial_x\delta(x-y)\,.
\end{multline}
By the way, we have shown that the Miura map (\ref{Miura}) is a kind of
Sugawara construction: it builds the stress-tensor-like object $u(x)$
out of the ``free scalar field" $q(x)$\,.

The first link between the KdV Poisson bracket (\ref{u-brac2})
and the Virasoro algebra is thus established. Let us try to elaborate
it further. Recall that the Virasoro algebra (Vir)
\begin{equation}
\label{Vir0}
[L_m,L_n] = (m-n)L_{m+n} + \frac{c}{12}\,m(m^2-1)\,\delta_{n,-m}
\end{equation}
can be realized as a centrally extended algebra of vector fields
$f(z)\,\partial_z$ on a circle,
\begin{equation}
\label{Vir2}
[f\partial,g\partial] \doteq (fg'-f'g)\,\partial
 + \frac{c}{12}\,\text{Res}_z (fg''')\,, \ \ \ \ \ \ \ f'\equiv\partial f\,,
\end{equation}
with $L_m=z^{-m+1}\partial_z$\,. Here $\text{Res}_z$ (coefficient of $z^{-1}$)
may be thought of as $\oint\!\frac{dz}{2\pi i}$\,, which we will denote
simply by $\int\!dz$\,. Note the property $\text{Res}_z f'\!=\!0$\,.
Now let us show that in terms of structure constants of the algebra
(\ref{Vir2})\,, the bracket (\ref{u-brac2}) assumes the form of Lie-Poisson
(Kirillov) bracket $\{\lambda_i,\lambda_j\}=c_{ij}^{k}\lambda_k$\,.
Viewing (\ref{Vir2}) as an expression for a component of the commutator
in a Lie algebra, like $[p,q]^k=c_{ij}^{k}p^i q^j$\,, we may cast it into
functional form, reserving a single discrete index 0 for central terms:
\begin{equation}
\label{funcVir}
[f,g](x) = \int\!dy\,dz\,C(x,y,z)f(y)g(z)\,, \ \ \ \ \ \
[f,g]_0 = \int\!dy\,dz\,C_0(y,z)f(y)g(z)\,,
\end{equation}
whence
\begin{equation}
\label{C}
C(x,y,z) = \delta(x-z)\,\partial_y\delta(y-z)
 - \delta(x-y)\,\partial_z\delta(y-z)\,, \ \ \ \ \ \
C_0(y,z) = - \frac{c}{12}\,\partial_z^3\delta(y-z)\,.
\end{equation}
Now a functional analog of the Lie-Poisson bracket arises:
\begin{equation}
\label{Kir}
\{\lambda(y),\lambda(z)\}
 = \lambda_0 C_0(y,z) + \int\!dx\,C(x,y,z)\lambda(x)
 = (\frac{\lambda_0 c}{12}\,\partial_y^3
 +(\lambda(y)+\lambda(z))\,\partial_y)\,\delta(y-z)\,.
\end{equation}
We assume $\lambda_0=1$\,, and observe that
the bracket (\ref{u-brac2}) may also be written as
\begin{equation}
\label{u-brac3}
\{u(x),u(y)\} = -(\frac{c}{12}\partial_x^3
 +(u(x)+u(y))\partial_x)\,\delta(x-y)
\end{equation}
owing to
\begin{equation}
\label{}
u(y)\partial_x\,\delta(x-y)
 = u(x)\partial_x\,\delta(x-y) + (y-x)u_x\,\partial_x\,\delta(x-y)
 = (u(x)\partial_x + u_x(x))\,\delta(x-y)\,.
\end{equation}
Therefore, it is of the Lie-Poisson form.

Lie-Poisson brackets are closely connected with the coadjoint action
of a Lie algebra on its dual. Let us now find a counterpart of the Poisson
structure $\Lambda_2$ (\ref{Lambda12}) in $\text{Vir}^*$\,, a dual space to
Vir. Proceeding a bit more accurately than above, we denote
a general element of the latter by a pair $(f,a)\equiv f\partial+ca$\,, with
$a$ being a number, and rewrite (\ref{Vir2}) as
\begin{equation}
\label{Vir3}
[(f,a),(g,b)] = (fg'-f'g\,,\ \frac{1}{12}\text{Res}_z\,fg''')\,.
\end{equation}
Let us also denote the adjoint action of an infinitesimal element
$(\varepsilon,\omega)$ (that is, a commutator
$[(\varepsilon,\omega),\,\cdot\ ]$) simply by $\delta_\varepsilon$\,. Then
\begin{equation}
\label{eps}
\text{ad}_{(\varepsilon,\omega)}(f,a)
 = (\delta_\varepsilon f,\delta_\varepsilon a)\,,
\ \ \ \ \ \delta_\varepsilon f = \varepsilon f' - f\varepsilon'
 = (-f\partial + f')\,\varepsilon,
\ \ \ \delta_\varepsilon a = \frac{1}{12}\text{Res}_z\, f'''\varepsilon.
\end{equation}
Further, let $(\lambda,\alpha) \in \text{Vir}^*$, \
$\alpha$ be a number, and the contraction be chosen in the form
\begin{equation}
\label{contr}
\langle(f,a)\,,(\lambda,\alpha)\rangle \doteq a\alpha
 + \text{Res}_z\,f\lambda\,.
\end{equation}
The invariance condition for (\ref{contr}) is
\begin{equation}
\label{invar}
\alpha\,\delta_\varepsilon a + a\,\delta_\varepsilon\alpha
 + \text{Res}_z\,(\lambda\,\delta_\varepsilon\!f
  + f\delta_\varepsilon\!\lambda)=0\,.
\end{equation}
This implies the following form of the coadjoint action
$\delta_\varepsilon(\lambda,\alpha)
\equiv \text{ad}_{(\varepsilon,\omega)}^{*}(\lambda,\alpha)$\,:
\begin{equation}
\label{epsdual}
\delta_\varepsilon\alpha = 0\,, \ \ \ \ \
\delta_\varepsilon\lambda = (\frac{\alpha}{12}\,\partial^3
 + 2\lambda\,\partial + \lambda')\,\varepsilon\,.
\end{equation}
Thus, $\Lambda_2$ (\ref{Lambda12}) corresponds (for $\alpha=6$) to
coadjoint action in Vir${}^*$.
This is in agreement with our previous observation that
the bracket (\ref{u-brac2}) is Lie-Poisson.

Moreover, eq.\,(\ref{epsdual}) provides additional evidence of an intimate
relation between the KdV solutions $u(x)$ and the energy-momentum tensor
$T(z)$ in CFT${}_2$\,. Recall the transformation rule of a primary field
$\varphi(z)$ of conformal dimension $\Delta$\,:
\begin{equation}
\label{prim}
\tilde{\varphi}(\tilde{z})(d\tilde{z})^\Delta = \varphi(z)(dz)^\Delta\,,
\ \ \ \ \ \
\tilde{z} = z + \varepsilon(z)\,.
\end{equation}
For infinitesimal $\varepsilon$\,, this reads
\begin{equation}
\label{eps2}
\tilde{\varphi}(z)-\varphi(z) \
 \simeq \ -(\Delta\varphi\partial + \varphi')\,\varepsilon\,.
\end{equation}
Analogous formulas for the stress tensor (which corresponds to $\Delta=2$
but is not a primary field due to an anomaly) look as follows:
\begin{equation}
\label{T'}
\tilde{T}(\tilde{z})(d\tilde{z})^2 = T(z)(dz)^2
 - \frac{c}{12}(dz)^2\,\{\tilde{z};z\}
\end{equation}
with $\{\tilde{z};z\}$ being the Schwarzian derivative
\begin{equation}
\label{Schwarz}
\{f(z)\,;z\} \doteq \frac{f'''}{f'}
 - \frac{3}{2}\left(\frac{f''}{f'}\right)^2
\end{equation}
with the properties
\begin{equation}
\label{}
\{t;z\}\,(dz)^2 = -\{z;t\}\,(dt)^2 = \{t;w\}\,(dw)^2 + \{w;z\}\,(dz)^2\,.
\end{equation}
In the infinitesimal form,
\begin{equation}
\label{epsT}
 \tilde{T}(z) - T(z) \
 \simeq \ -(\frac{c}{12}\partial^3 + 2T\partial + T')\,\varepsilon\,.
\end{equation}
Comparing $\delta_\varepsilon f$ in (\ref{eps}) with (\ref{eps2})\,,
and $\delta_\varepsilon\lambda$ in (\ref{epsdual}) with (\ref{epsT})\,,
we see that vector fields (elements of Vir) belong to $\Delta=-1$
whereas their duals to $\Delta=2$\,. Intrinsic duality of $\Delta$
and $1\!-\!\Delta$ spaces is well known in CFT${}_2$\,. It is due to
\begin{equation}
\label{dual}
\int\!dz\,f(z)\,\varphi(z) =
 \int\!f(z)(dz)^{1-\Delta}\,\varphi(z)(dz)^\Delta
\end{equation}
being a natural invariant contraction.

All the reasoning given above suggests that the KdV solutions
$u(x)$ also live in $\text{Vir}^*$ and so belong to the class $\Delta=2$
(with the central charge $c=6$)\,. Probably, the most direct evidence of
this fact is provided by the following observation. The Lax
operator~(\ref{L}) reveals the covariant properties generalizing
(\ref{prim}) with $\Delta=2$\,,
\begin{equation}
\label{covar}
\tilde{L}(\tilde{x}) = \partial_{\tilde{x}}^{2} + \tilde{u}(\tilde{x})
 = (\tilde{x}')^{-\frac{3}{2}}\,(\partial_x^2+u(x))\circ
  (\tilde{x}')^{-\frac{1}{2}}\,, \ \ \ \ \ \ \
\tilde{x}' \equiv \frac{d\tilde{x}}{dx}\,,
\end{equation}
if $u$ transforms abnormally, like a conformal stress tensor:
\begin{equation}
\label{u'}
\tilde{u}(\tilde{x})(d\tilde{x})^2 = u(x)(dx)^2
 - \frac{1}{2}(dx)^2\,\{\tilde{x};x\}\,.
\end{equation}
To show this, one uses
\begin{gather}
\label{}
\partial_{\tilde{x}}^{2} = \partial_{\tilde{x}}\circ\partial_{\tilde{x}}
 = (\tilde{x}')^{-1}\partial_x\circ(\tilde{x}')^{-1}\partial_x
 = (\tilde{x}')^{-2}\partial_x^2
  - \tilde{x}''(\tilde{x}')^{-3}\partial_x\,, \\
\partial_x^2\circ(\tilde{x}')^{-\frac{1}{2}}
 = -\frac{1}{2}\,(\tilde{x}')^{-\frac{1}{2}}\,\{\tilde{x};x\}
 - \tilde{x}''(\tilde{x}')^{-\frac{3}{2}}\partial_x
 + (\tilde{x}')^{-\frac{1}{2}}\,\partial_x^2\,.
\end{gather}
We conclude that not only the Poisson structure $\Lambda_2$\,,
but also the Lax operator of the KdV hierarchy, as well as all its
solutions, exhibit apparent features of conformal covariance, being
thus related with certain representations of the Virasoro algebra.

\end{document}